\documentclass[reprint, aps, pra, twocolumn, amsmath, amssymb, nofootinbib]{revtex4-1}

\usepackage[utf8]{inputenc}
\usepackage{amsfonts}
\usepackage{units}
\usepackage[normalem]{ulem}
\usepackage{graphicx}

\newcommand{\ecL}{\varepsilon_{\rm L}^{\rm c}}
\newcommand{\ecR}{\varepsilon_{\rm R}^{\rm c}}
\newcommand{\tL}{\mathcal{T}_{\rm L}}
\newcommand{\tR}{\mathcal{T}_{\rm R}}
\newcommand{\EF}{\varepsilon_{\rm F}}
\newcommand{\dmul}{\delta \mu_{\rm L}}
\newcommand{\dmur}{\delta \mu_{\rm R}}
\newcommand{\zj}{{\mathcal J}_0}
\newcommand{\zf}{{\mathcal F}_0}

\newcommand{\ipcms}{Universit\'e de Strasbourg, CNRS, Institut de Physique et Chimie des Mat\'eriaux de Strasbourg, UMR 7504, F-67000 Strasbourg, France}

\begin{document}

\title{Non-local effects in charge and energy transport with dissipative electrodes}

\author{Rodolfo A. Jalabert}
\affiliation{\ipcms}

\begin{abstract}
Recent advances in nano-thermometry motivate the extension of the Landauer-Büttiker scattering theory as to include the non-local dissipation associated with charge transport.  Such a program is implemented by describing the inelastic scattering in the connecting electrodes within an electrostatically self-consistent scheme.  The restriction to quasi-one-dimensional geometries,  weak excitation and low temperature allows to obtain general expressions of the current density and the dissipated power, valid in different regimes,  for the cases of an energy-independent mean-free-path or an energy-independent relaxation-rate.  In particular,  the dissipation asymmetry at both sides of a nano-device and the conditions for observing heating spots with a local maximum of the dissipated power are formulated in terms of the key parameters that define the nano-device and its environment.
\end{abstract}

\maketitle

\section{Introduction}

The question of where dissipation takes place in a setup of electronic transport through a quantum-coherent scatterer connected to electrodes has been debated since the formulation of the Landauer-Büttiker scattering theory of electrical resistance \cite{landauer1970,buttiker1986,landauer1987,Gurevich97,bringuier_hal2016,beenakker1991,datta1995,imry2002, jalabert2016}.  In such an approach,  each electrode is described as a macroscopic electron reservoir together with a quasi-one-dimensional lead supporting free electron propagation between the reservoir and the mesoscopic sample. Thus,  yielding the three basic components to which the description is reduced: scatterer,  reservoirs,  and  leads.  As a consequence,  dissipation has to occur at the reservoirs when the traversing electrons release their excess energy and thermalize.  The infinite size of the reservoirs makes dissipation possible,  and the perfect character of the leads accounts for the subtleties related with the connection between the classical and quantum realms. 

Recent experimental developments of local thermal probes in a variety of systems  \cite{tsutsui2012,lee2013,chapuis2015,menges2016,halbertal2016,halbertal2017,marguerite2019,
zeldov_pc2019,PhysRevApp2020,weng2021} allowed to enrich the conceptual discussion about dissipation with the information yielded by mappings of the local temperature and heating power in the neighborhood of an atomic or a nanoscale scatterer.  

The measured temperatures and the estimated heating,  obtained for atomic-scale junctions in the non-linear regime,  exhibited significant bias asymmetries \cite{tsutsui2012,lee2013},  with more dissipation downstream than upstream of the scatterer in cases with a strongly energy-dependent transmission.  Combining the scattering approach with ab initio calculations for molecular junctions resulted in a heating which depends on the current direction and that is not equally splitted between the two electrodes feeding the junction \cite{zotti2014},  consitently with the experimental findings.  The incorporation of realistic effects in the theory,  such as inelastic and dephasing processes,  was later shown to reduce such an asymmetry in the nonlinear transport regime \cite{sanchez2015}. 

Scanning thermal microscopy allowed to resolve temperature variations of the order of milli-Kelvins with nanoscale spatial resolution in different physical systems (two-dimensional electron gases of a semi-conductor hetero-structure or a graphene sheet,  as well as  carbon nanotubes) \cite{halbertal2016,halbertal2017,marguerite2019} showing a remarkable spatial separation between the place where the voltage drops and that where the associated heating occurs.  In particular,  for a quantum-point-contact connected to two-dimensional electrodes,  important asymmetries of the heating are observed,  with well-defined hot spots ({\it i.e.}~maxima of the local temperature) downstream of the nano-structure acting as a scatterer \cite{zeldov_pc2019}.  Such a non-local dissipation effect was in line with the theoretical prediction \cite{rokni1995} for a metallic quantum point contact (having an opening much larger than the elastic mean-free path) of a separation between the potential drop and the delivery of the Joule heat. 

The application of the scattering approach in the ballistic regime \cite{SciPostPhys2022} resulted in an asymmetry of the dissipation at both sides of a scatterer,  with larger heating downstream in the direction of the carriers  when the transmission coefficient monotonically  increases with energy,  as is the case of a conductance step for a quantum-point-contact.  While the scattering approach by itself does not allow to obtain the position of hot spots,  an estimation for the location of the maximum dissipation could be established by assimilating the two-dimensional electron gases at both sides of the quantum-point-contact to electron reservoirs.  The electron-electron and the electron-phonon interactions in the two-dimensional electron gas are responsible for a finite lifetime of Landau quasi-particles according to their excess energy \cite{jalabert1989},  thus governing the distance over which thermalization occurs \cite{SciPostPhys2022}.  
 
Outside the scattering approach,  an asymmetric heat dissipation was obtained,   in the hydrodynamic regime,  for a disordered quasi-one-dimensional geometry with a spatially inhomogeneous distribution of the relaxation-rate \cite{Mirlin2019}.  Such an non-homogeneity mimics the case of a potential barrier as a scatterer,  but it ignores its quantum coherence by describing the whole system within the Boltzmann equation.

Based on previous attempts to generalize the Landauer conductance formula \cite{eranen1987, laikhtman1994,landauer1995},   Ref.~\cite{NicoPRB} considered a model-system of a coherent scatterer embedded in a one-dimensional wire where relaxation takes place,  by treating the scatterer within the scattering approach and describing the inelastic electron-phonon relaxation in the wires through the Boltzmann equation. The strong Coulomb limit ensuring local charge neutrality,  and a velocity-independent relaxation-rate for the wires,  allowed to establish the power dissipation asymmetry and the existence of thermal spots.  The latter,  given by a maximum or a minimum of the dissipated power density (corresponding,  respectively,  to a heating or a cooling spot),  occur  at a distance from the scatterer of the order of the electron-phonon relaxation length,  according to the values taken by the various parameters of the model.  An exhaustive numerical search determined that the nature and the conditions for the appearance of thermal spots at each side of the scatterer are governed by the temperature and the imposed excitation,  as well as by the precise energy-dependence of the transmission through the scatterer and of the relaxation-rate in the wires.  In particular,  scatterers exhibiting a sharp energy-dependent transmission coefficient ({\it i.e.} a strongly peaked Lorentzian acting as an energy-filter),  were shown to lead,  at relatively high temperatures and excitation voltages,  to the ubiquitous appearance of heating and cooling spots.

In order to advance our understanding of the experimentally found heat dissipation asymmetry and hot spots,  we develop a theoretical approach to the problem of a coherent scatterer connected to dissipative wires,  determining the role of some crucial aspects of the problem,  like the energy-dependence of the scatterer transmission coefficient and of the mean-free-path in the wires,  as well as Coulomb and temperature effects.  The particular cases of a velocity-independent mean-free-path \cite{eranen1987} and a velocity-independent relaxation-time \cite{NicoPRB}  allow for considerable analytical progress in the weak-excitation,  low-temperature regime,  leading to a generalized Landauer formula and the main features of the power dissipation. 

The paper is organized as follows. Section \ref{sec:QODM} presents the model system,  focusing in the cases of a velocity-independent mean-free-path and a velocity-independent relaxation-time,  the latter only treated in the weak-excitation,  low-temperature regime.   Section \ref{sec:dp} provides the general features of the dissipated power and applies them to the previous two cases,  determining the dissipation asymmetry and discussing the existence of thermal spots.  We provide the conclusions of our work in Sec. ~\ref{sec:conclusions}.  We relegate to the appendices technical aspects of our work: the self-consistent treatment of the electrostatic potential 
(App.  \ref{sec:appendixSCTEP}),  the weak-excitation low-temperature regime for the cases of a velocity-independent mean-free-path (App. \ref{sec:Appltlv}) or velocity-independent relaxation-time 
(App. \ref{sec:AppWELTCRL}),  as well as the results for the strong Coulomb limit of vanishing screening length for the two cases studied (Apps.  \ref{sec:scl} and \ref{sec:sclcrt}).

\vspace{0.5cm}

\section{Quasi-one dimensional model}
\label{sec:QODM}

Adopting the model of Refs.~\cite{eranen1987} and \cite{NicoPRB},  we consider carriers of charge $e$ in a wire of cross-section $A$,  directed along the $z$-axis,  with a quantum-coherent scatterer (typically a tunnel barrier) placed around the origin (within the interval $[z_{-}, z_{+}]$, with $z_{-}<0$ and $z_{+}>0$,  $L=z_{+}-z_{-}$).  The scatterer is characterized by an energy-dependent transmission coefficient $\mathcal{T}(\varepsilon)$,  and the scattering within the wire is described by a mean-free-path $l(v)$ that,  in the generic case,  depends on the carrier velocity $v$.  Working within the Boltzmann equation and postulating a relaxation-time approximation to a local-equilibrium distribution \cite{AM},  lead to space- and velocity-dependent non-equlibrium distribution functions for the carriers,   which,  at the left (L) and at the right (R) of the scatterer,  can be,  respectively,  written as \cite{eranen1987}

\begin{figure}[h!]
  \includegraphics[width=\linewidth]{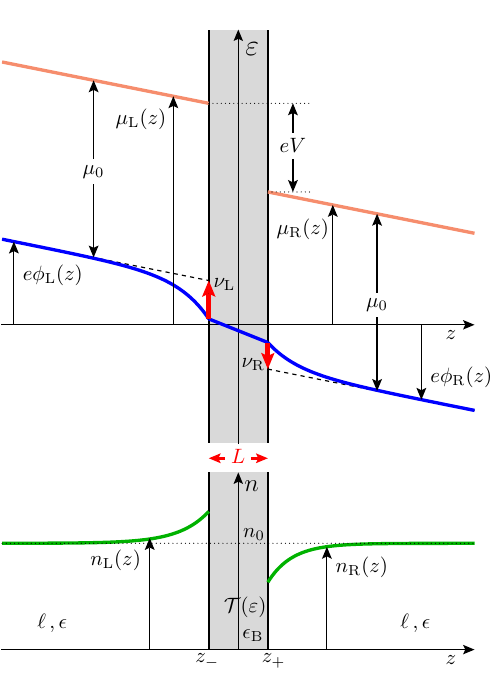}
       \caption{
       Energy diagram (upper panel) and carrier density (lower panel) for the model-system of a scatterer (shadowed region of length $L$ defined by $z_{-} \le z \le z_{+}$) connected to two semi-infinite wires.  The scatterer (assimilated to a tunnel barrier) is characterized by an energy-dependent transmission coefficient $\mathcal{T}(\varepsilon)$ and a permittivity $\epsilon_{\rm B}$,  the wires are characterized by a velocity-independent mean-free-path $\ell$ and permittivity $\epsilon$.   The parameters $\nu_{\rm L,R}$ are defined in Eq.~\eqref{eq:nuLR},  $\mu_0$ is the equilibrium chemical potential found in the wires far away from the scatterer,  $n_0$ is the equilibrium carrier density,  $e$ is the charge of the carriers,  and $V$ is the potential drop due to the presence of the scatterer.  The electro-chemical potential $\mu_{\rm L,R}(z)$,  the electrostatic potential energy $e\phi_{\rm L,R}(z)$,  and the carrier density $n_{\rm L,R}(z)$ for the left (right) wire are labeled by arrows and represented by the orange,  blue,  and green curves,  respectively.  The linear drop of the electrostatic potential across the scatterer follows from the assumption that the scatterer is electrically  neutral.  The linear behavior of the $\mu_{\rm L,R}(z)$ in the wires is a consequence of the assumption of a velocity-independent mean-free-path.  For the sake of presentation, the non-equilibrium features are exaggerated.}
    \label{fig:setup}
\end{figure}

\begin{widetext}
\begin{subequations}
\label{eq:distributions}
\begin{align}
\label{eq:distributionLp}
g_{\rm L}(z,v_{\rm L}) & = f(\ecL) - 
f^{\prime}(\ecL)  
\left[ \delta \mu_{\rm L}(z) - e \phi_{\rm L}(z) \right] +
f^{\prime}(\ecL) \int_{-\infty}^{z} {\rm d}z^{\prime} \ 
\dmul^{\prime}(z') \ \exp{\left(\frac{z^{\prime}-z}{l(v_{\rm L})} \right)}
 \,  ,
\\
\label{eq:distributionLn}
g_{\rm L}(z,-v_{\rm L}) & = f(\ecL) - 
f^{\prime}(\ecL)  \left[ \delta \mu_{\rm L}(z) - e \phi_{\rm L}(z) \right] -
f^{\prime}(\ecL) \int_{z}^{z_{-}} {\rm d}z^{\prime} \
\dmul^{\prime}(z') \
\exp{\left(\frac{z-z^{\prime}}{l(v_{\rm L})} \right)} + 
g_{\rm L}^{(1)}(-v_{\rm L}) \ \exp{\left(\frac{z-z_{-}}{l(v_{\rm L})} \right)} 
\,  ,
\\
\label{eq:distributionRp} 
g_{\rm R}(z,v_{\rm R}) & = f(\ecR) - 
f^{\prime}(\ecR)  \left[ \delta \mu_{\rm R}(z) - e \phi_{\rm R}(z) \right] +
f^{\prime}(\ecR) \int_{z_{+}}^{z} {\rm d}z^{\prime}  \
\dmur^{\prime}(z') \
\exp{\left(\frac{z^{\prime}-z}{l(v_{\rm R})} \right)} + 
g_{\rm R}^{(1)}(v_{\rm R}) \ \exp{\left(\frac{z_{+}-z}{l(v_{\rm R})} \right)} 
 \,  ,
 \\
\label{eq:distributionRn} 
g_{\rm R}(z,-v_{\rm R}) & = f(\ecR) - 
f^{\prime}(\ecR)  \left[ \delta \mu_{\rm R}(z) - e \phi_{\rm R}(z) \right] -
f^{\prime}(\ecR) \int_{z}^{\infty} {\rm d}z^{\prime} \
\dmur^{\prime}(z') \
\exp{\left(\frac{z-z^{\prime}}{l(v_{\rm R})} \right)} 
 \,  ,
\end{align}
\end{subequations}

\end{widetext}
for right (left) movers (we only consider $v_{\rm L,R}>0$). 
The kinetic energies at the left (right) of the scatterer are $\varepsilon_{\rm L,R}^{\rm c}=mv_{\rm L,R}^2/2$.  Since the scattering is elastic,  for a traversing electron we have $\ecR=\ecL+eV$,  with $V$ the potential drop induced by the scatterer.  We note $f$ the Fermi distribution function,  {\it i.e},
\begin{equation}
f(\varepsilon)=\frac{1}{\exp{\left(\left[\varepsilon-\mu_{0}\right]/k_{\rm B}T\right)}+1} \,  ,
\end{equation}
and $f^{\prime}(\varepsilon)$ its energy-derivative,  with $\mu_{0}$ the equilibrium chemical potential,  taken to be the same at both sides of the scatterer since the contacts are chosen of the same material.  At the left (right) of the scatterer,  the local electro-chemical potential is $\mu_{\rm L,R}(z)=\mu_{0}+\delta \mu_{\rm L,R}(z)$,  and $\phi_{\rm L,R}(z)$ is the electrostatic potential (see the upper panel of Fig.~\ref{fig:setup} for the energy diagram in the case of a velocity-independent mean-free-path $\ell$).  We denote $\mu_{\rm L,R}^{\prime}(z)$,  $\delta \mu_{\rm L,R}^{\prime}(z)$,  and $\phi_{\rm L,R}^{\prime}(z)$ the $z$-derivative of the above-defined functions.  The first two terms in each of Eqs.~\eqref{eq:distributions} represent a linearization of the local-equilibrium distribution 
$g_{\rm L,R}^{(\rm le)}(z,v_{\rm L,R}) = f(\varepsilon_{\rm L,R}^{\rm c}+e \phi_{\rm L,R}(z)-\delta \mu_{\rm L,R}(z))$,  which does not depend on the direction of the velocity. The remaining terms in each of Eqs.~\eqref{eq:distributions} stand for the modification of the local-equilibrium distribution in the left (right) wire,  set by the field $-e \phi_{\rm L,R}^{\prime}(z)$ and the boundary conditions imposed by the scatterer.  

The scatterer is characterized by its energy-dependent transmission coefficient $\mathcal{T}(\varepsilon)$. We choose the energy origin such that $\varepsilon=\varepsilon_{\rm L,R}^{\rm c} \pm eV/2$. The matching conditions across the scatterer can be expressed as
\begin{widetext}
\begin{subequations}
\label{eq:matchingconditions}
\begin{align}
\label{eq:matchingconditionLg}
g_{\rm L}(z_{-},-v_{\rm L}) & = g_{\rm L}(z_{-},v_{\rm L})+ \mathcal{T}(\varepsilon) \left[g_{\rm R}(z_{+},-v_{\rm R}) -
g_{\rm L}(z_{-},v_{\rm L}) \right]
 \,  ,
\\
\label{eq:matchingconditionRg}
g_{\rm R}(z_{+},v_{\rm R}) & = g_{\rm R}(z_{+},-v_{\rm R})+ \mathcal{T}(\varepsilon) \left[g_{\rm L}(z_{-},v_{\rm L}) -
g_{\rm R}(z_{+},-v_{\rm R}) \right]
 \,  ,
\end{align}
\end{subequations}
and allow to write the space-independent factors $g_{\rm L}^{(1)}(-v_{\rm L})$ and $g_{\rm R}^{(1)}(v_{\rm R})$ of 
Eqs. \eqref{eq:distributions} as 
\begin{subequations}
\label{eq:g1Landg1R}
\begin{align}
\label{eq:g1L}
g_{\rm L}^{(1)}(-v_{\rm L}) & =
f^{\prime}(\ecL) \int_{-\infty}^{z_{-}} {\rm d}z  \
\dmul^{\prime}(z) \
\exp{\left(\frac{z-z_{-}}{l(v_{\rm L})} \right)}
+ \mathcal{T}(\varepsilon)
\Biggl(
f(\ecR) - f(\ecL) + \nu_{\rm L} \ f^{\prime}(\ecL) -
\nu_{\rm R} \ f^{\prime}(\ecR)    \nonumber \\
&
\hspace*{0.5cm}
- f^{\prime}(\ecL) \int_{-\infty}^{z_{-}} {\rm d}z  \
\dmul^{\prime}(z) \
\exp{\left(\frac{z-z_{-}}{l(v_{\rm L})} \right)}
 -
f^{\prime}(\ecR) \int_{z_{+}}^{\infty} {\rm d}z \
\dmur^{\prime}(z) \
\exp{\left(\frac{z_{+}-z}{l(v_{\rm R})} \right)} 
\Biggr)
 \,  ,
\\
\label{eq:g1R}
g_{\rm R}^{(1)}(v_{\rm R}) & = - f^{\prime}(\ecR)
\int_{z_{+}}^{\infty} {\rm d}z  \
\dmur^{\prime}(z) \
\exp{\left(\frac{z_{+}-z}{l(v_{\rm R})} \right)}
+ \mathcal{T}(\varepsilon)
\Biggl( 
f(\ecL) - f(\ecR) - \nu_{\rm L} \ f^{\prime}(\ecL) +
\nu_{\rm R} \ f^{\prime}(\ecR)
  \nonumber \\
&
\hspace*{0.5cm}
+
 f^{\prime}(\ecL)  \int_{-\infty}^{z_{-}} {\rm d}z \
\dmul^{\prime}(z) \
\exp{\left(\frac{z-z_{-}}{l(v_{\rm L})} \right)} 
 + f^{\prime}(\ecR) \int_{z_{+}}^{\infty} {\rm d}z \
\dmur^{\prime}(z) \
\exp{\left(\frac{z_{+}-z}{l(v_{\rm R})} \right)} 
\Biggr)
 \,  ,
\end{align}
\end{subequations} 
\end{widetext}
where we have defined the parameters
\begin{equation}
\label{eq:nuLR}
\nu_{\rm L,R}=\delta \mu_{\rm L,R}(z_{\mp}) - e \phi_{\rm L,R}(z_{\mp}) \,  .
\end{equation}
The carrier density (including the spin-degeneracy factor) is
%
%
\begin{align}
\label{eq:ndef}
n_{\rm L,R}(z) & =  \frac{m}{\pi A \hbar} \int_{0}^{\infty} {\rm d}v_{\rm L,R} \left[g_{\rm L,R}(z,v_{\rm L,R})+g_{\rm L,R}(z,-v_{\rm L,R}) 
\right] 
\nonumber \\
& = n^{(0)} + \delta n_{\rm L,R}(z)
\,  ,
\end{align}
with the equilibrium density 
\begin{equation}
\label{eq:n0def}
n^{(0)} = \frac{2 m}{\pi A \hbar} \int_{0}^{\infty} {\rm d}v_{\rm L,R} \
f(\varepsilon_{\rm L,R}^{\rm c})
 \,  ,
\end{equation}
and its non-equilibrium correction $\delta n_{\rm L,R}(z)$,  obtained by using the expressions \eqref{eq:distributions} and \eqref{eq:g1Landg1R} into Eq.~\eqref{eq:ndef}.  At this point,  it is useful to also introduce the local-equilibrium density 
\begin{equation}
\label{eq:nledef}
n^{\rm (le)}_{\rm L,R}(z) = \frac{2 m}{\pi A \hbar} \int_{0}^{\infty} {\rm d}v_{\rm L,R} \ g_{\rm L,R}^{({\rm le})}(z,v_{\rm L,R})
 \,  .
\end{equation}
The Fermi form of the functions $f(\varepsilon)$ and 
$g_{\rm L,R}^{({\rm le})}(z,v_{\rm L,R})$ results in
\begin{equation}
\label{eq:n0exact}
n^{(0)} = - \frac{1}{A \hbar} \sqrt{\frac{2 m k_{\rm B}T}{\pi}} \
{\rm Li}_{1/2}\Big(\!-\exp{\left(\mu_{0}/k_{\rm B}T\right)} \Big)
 \,  ,
\end{equation}
and $n^{\rm (le)}_{\rm L,R}(z)$ given by the right-hand-side of the above expression,  up to the substitution of $\mu_{0}$ by 
$\mu_{\rm L,R}(z)-e\phi_{\rm L,R}(z)$.  We note ${\rm Li}_{k}(x)$ the polylogarithmic function of order $k$.  Within the adopted linearization,  
\begin{equation}
\label{eq:nle}
n^{\rm (le)}_{\rm L,R}(z) =  n^{(0)} + \delta n^{\rm (le)}_{\rm L,R}(z)
\,  ,
\end{equation}
with
\begin{equation}
\label{eq:deltanle}
\delta n^{\rm (le)}_{\rm L,R}(z) = -2 \ {\cal F}_0  
\Bigl[
\delta \mu_{\rm L,R}(z) - e \phi_{\rm L,R}(z) 
\Bigr]
\,  ,
\end{equation}
and
\begin{equation}
\label{eq:f0}
{\cal F}_0
 = \frac{m}{\pi A \hbar} \int_{0}^{\infty} {\rm d}v_{\rm L,R} \ f^{\prime}(\varepsilon_{\rm L,R}^{\rm c})
 \,  .
\end{equation}

The accumulated charge at the left (right) of the scatterer follows from the carrier excess at both sides,
\begin{subequations}
\label{eq:accumulated charge}
\begin{align}
{\cal N}_{\rm L} & =  \int_{-\infty}^{z_{-}} {\rm d}z \
\delta n_{\rm L}(z) 
 \,  ,
\\
{\cal N}_{\rm R} & =  \int_{z_{+}}^{\infty} {\rm d}z \
\delta n_{\rm R}(z) 
 \,  ,
\end{align}
\end{subequations}
defining the residual resistivity dipole (also called Landauer dipole) \cite{landauer1975,landauer1987,datta1995,sorbello1989,zwerger1991,
datta91,laikhtman1994,kunze1995,nanolett2025}.  The global charge neutrality  is assured by the condition ${\cal N}_{\rm L}+{\cal N}_{\rm R}= 0$ (see the lower panel of Fig.~\ref{fig:setup} for a sketch of the carrier density in the case of a velocity-independent mean-free-path $\ell$).  

The form \eqref{eq:distributions} of the non-equilibrium distribution functions implies the useful identity
\begin{widetext}
\begin{equation}
\label{eq:useful}
\frac{\partial}{\partial z} \
\left[ 
g_{\rm L,R}(z,v_{\rm L,R}) - g_{\rm L,R}(z,-v_{\rm L,R})
\right]  = -
\frac{1}{l(v_{\rm L,R})} \
\left[ 
g_{\rm L,R}(z,v_{\rm L,R}) + g_{\rm L,R}(z,-v_{\rm L,R}) - 
2 \ g_{\rm L,R}^{({\rm le})}(z,v_{\rm L,R})
\right] 
 \,  .
\end{equation}

The current-conserving condition allows to express the current density $j$ under alternative forms,  that we write (including the spin-degeneracy factor) as 
\begin{subequations}
\label{eq:jdef}
\begin{align}
\label{eq:jdefa}
j = \frac{em}{\pi A \hbar} \int_{0}^{\infty} {\rm d}v_{\rm L,R} \ 
v_{\rm L,R} \left[ g_{\rm L,R}(z,v_{\rm L,R}) - g_{\rm L,R}(z,-v_{\rm L,R}) \right] 
 \,  ,
\\
\label{eq:jdefb}
j =
 \frac{em}{\pi A \hbar} \int_{0}^{\infty} {\rm d}v_{\rm L,R} \
v_{\rm L,R} \ \mathcal{T}(\varepsilon)
\left[ g_{\rm L}(z_{-},v_{\rm L}) - g_{\rm R}(z_{+},-v_{\rm R}) \right] \,  .
\end{align}
\end{subequations}

In the steady-state regime that we work with, $j$ is $z$-independent,  and therefore,  using \eqref{eq:useful} and \eqref{eq:jdefa} we obtain
%
\begin{equation}
\label{eq:pccgen}
\int_{0}^{\infty} {\rm d}v_{\rm L,R} \ 
\frac{v_{\rm L,R}}{l(v_{\rm L,R})} \
\left[ 
g_{\rm L,R}(z,v_{\rm L,R}) + g_{\rm L,R}(z,-v_{\rm L,R}) - 
2 \ g_{\rm L,R}^{(0)}(z,v_{\rm L,R})
\right]  = 0 \,  ,
\end{equation}
which expresses the condition of the particle conservation in the local relaxation.

Using the $v_{\rm L}$-integral in Eq.~\eqref{eq:jdefa},  together with the expressions \eqref{eq:distributions} and \eqref{eq:g1Landg1R},  we have
\begin{align}
\label{eq:jsc}
j & = \frac{em}{\pi A \hbar} \int_{0}^{\infty} {\rm d}v_{\rm L} \ 
v_{\rm L} 
\Biggl[
f^{\prime}(\ecL) \int_{-\infty}^{z} {\rm d}z^{\prime} \
\dmul^{\prime}(z') \
\exp{\left(\frac{z^{\prime}-z}{l(v_{\rm L})} \right)} 
 +
f^{\prime}(\ecL) \int_{z}^{z_{-}} {\rm d}z^{\prime}  \
\dmul^{\prime}(z') \
\exp{\left(\frac{z-z^{\prime}}{l(v_{\rm L})} \right)}
\nonumber \\
& 
- \exp{\left(\frac{z-z_{-}}{l(v_{\rm L})} \right)}
\left\{
f^{\prime}(\ecL) \int_{-\infty}^{z_{-}} {\rm d}z'  \
\dmul^{\prime}(z') \
\exp{\left(\frac{z'-z_{-}}{l(v_{\rm L})} \right)}
+ \mathcal{T}(\varepsilon)
\Biggl(
f(\ecR) - f(\ecL) + \nu_{\rm L} \ f^{\prime}(\ecL) -
\nu_{\rm R} \ f^{\prime}(\ecR) \right.  
\nonumber \\
&
\left.
- f^{\prime}(\ecL) \int_{-\infty}^{z_{-}} {\rm d}z' \
\dmul^{\prime}(z') \
\exp{\left(\frac{z'-z_{-}}{l(v_{\rm L})} \right)} 
 -
f^{\prime}(\ecR) \int_{z_{+}}^{\infty} {\rm d}z'  \
\dmur^{\prime}(z') \
\exp{\left(\frac{z_{+}-z'}{l(v_{\rm R})} \right)}
\Biggr)
\right\}
\Biggr] 
 \,  .
 \end{align}
A similar relation (not shown) can be obtained for $z > z_{+}$ by using the $v_{\rm R}$-integral in Eq.~\eqref{eq:jdefa},  thus  completing a system of coupled integral equations in $\dmul^{\prime}(z)$ and $\dmur^{\prime}(z)$,  which is quite difficult to treat in the general case.  However,  considerable progress can be made in the special cases of a velocity-independent mean-free-path $\ell$  \cite{eranen1987} or a velocity-independent relaxation time $\tau$.  We treat each of these cases in the sequel. 

\subsection{Velocity-independent mean-free-path $\ell$}
\label{subsec:cmfp}

A velocity-independent  mean-free-path $l \equiv \ell$ allows to decouple the system of integral equations in $\delta \mu_{\rm L,R}^{\prime}(z)$ and,  once the condition that $j$ is $z$-independent is implemented,  Eq.~\eqref{eq:jsc} results in
\begin{align}
\label{eq:jsccmfp}
  \int_{-\infty}^{z} {\rm d}z^{\prime} \ &  \
\dmul^{\prime}(z') \
\exp{\left(\frac{z^{\prime}-z}{\ell} \right)} -
 \int_{z}^{z_{-}} {\rm d}z^{\prime}  \
\dmul^{\prime}(z') \
\exp{\left(\frac{z-z^{\prime}}{\ell} \right)}
\nonumber \\
& 
\hspace*{0.5cm}
+
\exp{\left(\frac{z-z_{-}}{\ell} \right)}
\left\{ 
\int_{-\infty}^{z_{-}} {\rm d}z' \
\dmul^{\prime}(z') \
\exp{\left(\frac{z'-z_{-}}{\ell} \right)} 
- \frac{j}{\zj}
\right\} = 0
 \,  ,
 \end{align}
%
with
\begin{align}
\label{eq:J0}
{\mathcal J}_0 & = \frac{em}{\pi A \hbar} \int_{0}^{\infty} {\rm d}v_{\rm L,R} \ 
v_{\rm L,R} \ f^{\prime}(\varepsilon_{\rm L,R}^{\rm c}) 
= - \frac{e}{2 \pi A \hbar} \left[1 + \tanh{\left(\frac{\mu_{0}}{2k_{\rm B}T}\right)} \right]
\, .
\end{align}
\end{widetext}

The solution for Eq.~\eqref{eq:jsccmfp},  as well as that of the analogous equation for $\dmur^{\prime}(z)$,  have the simple form \cite{eranen1987}
\begin{equation}
\label{eq:chpotcmfp}
\delta \mu_{\rm L,R}^{\prime}(z) =
 \frac{j}{2 {\mathcal J}_0 \ell} \,  ,
\end{equation}
indicating that the linear drop of the electro-chemical potential characteristic of a Drude wire with a mean-free-path $\ell$ is not affected by the presence of the scatterer,  as shown in Fig.~\ref{fig:setup}.  This important result translates in simplified expressions of the space-independent factors $g_{\rm L,R}^{(1)}(\mp v_{\rm L,R})$ (defined in Eqs.~\eqref{eq:g1Landg1R}),  the non-equilibrium correction of the carrier density $\delta n_{\rm L,R}(z)$ (defined in Eq.~\eqref{eq:ndef}),  and the particle-conserving condition \eqref{eq:pccgen},  that can be written,  respectively,  as 
%
\begin{align}
\label{eq:g1LRsim}
&g_{\rm L,R}^{(1)}(\mp v_{\rm L,R})  = 
\pm f^{\prime}(\varepsilon_{\rm L,R}^{\rm c}) \frac{j}{2 \zj}
\mp \mathcal{T}(\varepsilon) 
\Biggl(
f(\ecL) - f(\ecR) 
\nonumber
\\
- & 
\nu_{\rm L} \ f^{\prime}(\ecL) +
\nu_{\rm R} \ f^{\prime}(\ecR) +
\Bigl(f^{\prime}(\ecL) + f^{\prime}(\ecR)  \Bigr)
\frac{j}{2 \zj} \Biggr)
 \,  , 
\end{align}
%
\begin{align}
\label{eq:deltan}
\delta n_{\rm L,R}(z)  =  & -2 \ {\cal F}_0  
\Bigl[
\delta \mu_{\rm L,R}(z) - e \phi_{\rm L,R}(z) 
\Bigr] 
\nonumber \\ 
&
\pm q_{\rm L,R} \ \exp{\left(\pm \ \frac{z-z_{\mp}}{\ell} \right)} 
 \,  ,
\end{align}
and
\begin{equation}
\label{eq:pcc}
\int_{0}^{\infty} {\rm d}v_{\rm L,R} \ 
v_{\rm L,R} \left[ g_{\rm L,R}^{(1)}(\mp v_{\rm L,R}) \pm \frac{j}{2 \zj} f^{\prime}(\varepsilon_{\rm L,R}^{\rm c}) \right]  = 0 \,  .
\end{equation}
In Eq.~\eqref{eq:deltan} we have defined
\begin{subequations}
\label{eq:qLR0}
\begin{align}
q_{\rm L}   & = \alpha +  \mathcal{T}_{\rm L}^{\rm L} \ \nu_{\rm L} -  \mathcal{T}_{\rm L}^{\rm R} \ \nu_{\rm R}
 \,  ,
\\
q_{\rm R}   & = \beta +  \mathcal{T}_{\rm R}^{\rm L} \ \nu_{\rm L} -  \mathcal{T}_{\rm R}^{\rm R} \ \nu_{\rm R} 
 \,  ,
\end{align}
\end{subequations}
with
%
\begin{subequations}
\label{eq:defctes}
\begin{align}
\label{eq:alphadef}
\alpha  = & \frac{j}{\zj} 
\left(
 \zf - \frac{1}{2}
 \left[
 \mathcal{T}_{\rm L}^{\rm L} + 
 \mathcal{T}_{\rm L}^{\rm R}
\right]
 \right) 
 \nonumber \\ & 
 - 
 \frac{m}{\pi A \hbar} \int_{0}^{\infty} {\rm d}v_{\rm L} \ 
\mathcal{T}(\varepsilon) 
\left[ f(\ecL) - f(\ecR) \right]
 \,  ,
\\
\label{eq:betadef}
\beta  = & \frac{j}{\zj} 
\left(
\zf - \frac{1}{2}
 \left[
 \mathcal{T}_{\rm R}^{\rm R} + 
\mathcal{T}_{\rm R}^{\rm L}
\right]
 \right) 
 \nonumber \\ & 
- 
 \frac{m}{\pi A \hbar} \int_{0}^{\infty} {\rm d}v_{\rm R} \ 
\mathcal{T}(\varepsilon) 
\left[ f(\ecL) - f(\ecR) \right]
 \,  ,
\end{align}
\end{subequations}
and
%
\begin{equation}
\label{eq:expansctestauLRLR}
 \mathcal{T}_{\rm k}^{\rm p}  = \frac{m}{\pi A \hbar} 
\int_{0}^{\infty} {\rm d}v_{\rm k} \ \mathcal{T}(\varepsilon) \ f^{\prime}(\varepsilon_{\rm p}^{\rm c}) \,  ,  \hspace*{0.1cm}  {\rm for} \, \,
{\rm k,p}= {\rm L,R}
\,  .
\end{equation}
In addition,  the form \eqref{eq:jdefa} of the current density can be expressed as \cite{eranen1987}
\begin{widetext}
\begin{equation}
\label{eq:j4p}
j = \frac{1}{1-\frac{1}{2} \left(\tL + \tR \right)} \left\{  \frac{em}{\pi A \hbar} \int_{0}^{\infty} {\rm d}v_{\rm L} \ v_{\rm L} \
\mathcal{T}(\varepsilon) 
\Big[ f(\ecL) - f(\ecR)  \Big] - \zj \ \nu_{\rm L} \ \tL +
\zj \ \nu_{\rm R} \ \tR \right\} \,  ,
\end{equation}
\end{widetext} 
with
\begin{equation}
\label{eq:T_LR}
 \mathcal{T}_{\rm L,R} = \frac{1}{\zj} \ \frac{em}{\pi A \hbar}
 \int_{0}^{\infty} {\rm d}v_{\rm L,R} \ v_{\rm L,R } \ \mathcal{T}(\varepsilon) \ f_{0}^{\prime}(\varepsilon_{\rm L,R}^{\rm c}) \,  .
\end{equation}
%
Eq.~\eqref{eq:j4p} has the structure of the 4-point Landauer formula,  to which it reduces in the trivial limit of $\ell \rightarrow \infty$ and non-interacting carriers (for which $\nu_{\rm L,R}=0$),  when considering the scattering between left and right reservoirs with electro-chemical potentials $\mu_{\rm L,R} = \mu_0 \pm e V/2$.  Outside this limiting case,  Eq.~\eqref{eq:j4p} is an implicit current-voltage characteristics,  since $\nu_{\rm L,R}$ depend on $j$,  and the solution of the electrostatic problem has to be implemented in order to determine the relationship between these last two physical quantities.  Such an approach is undertaken in Appendix \ref{sec:appendixSCTEP},  where the explicit current-voltage characteristics \eqref{eq:j4pex} is obtained.  

\subsubsection{Weak-excitation,  low-temperature regime}
\label{subsec:ltlcrt}

Working in the weak-excitation,  low-temperature (WELT) regime under the assumption of a smooth energy-dependence of $\mathcal{T}(\varepsilon)$,  as done in Appendix \ref{sec:Appltlv},  allows to obtain simpler forms of the parameters \eqref{eq:f0},  \eqref{eq:J0},  \eqref{eq:qLR0}-\eqref{eq:expansctestauLRLR},  and \eqref{eq:T_LR},  as well as other  physical quantities of interest.  In particular the explicit current-voltage characteristics \eqref{eq:j4pex} takes the form 
\begin{widetext}
\begin{equation}
\label{eq:j4plT}
j =  \frac{G_0}{A} \ V
\left( \frac{\mathcal{\tilde T}_0}{1-\mathcal{\tilde T}_0} \right) 
\left(
\frac{1+2\epsilon_{\rm B}/\omega}{1+\tilde{\kappa} } 
\right)
\left(1 -  d_{\rm T}
\
\frac{ \kappa}{1+\kappa} \
 \frac{\left( k_{\rm s} \ell \right)^2 }{k_{\rm s}\ell+1}
\right)
\,  ,
\end{equation}
\end{widetext}
with $G_0=e^2/\pi \hbar$ the conductance quantum.  As in Appendix \ref{sec:Appltlv},  we use the sub-index 0 for the value of the energy-dependent functions at $\mu_0$ ({\it i.e.} ${\mathcal T}_0 = {\mathcal T}(\mu_0)$),  while the renormalized transmission coefficient $\tilde{\mathcal T}(\varepsilon)$ is given in Eq.~\eqref{eq:tautilde}.  We have defined the thermal factor
\begin{equation}
\label{eq:therfac}
d_{\rm T} = 
\frac{\pi^2}{3}  \frac{\left(k_{\rm B}T\right)^2}{\mu_0} \
\frac{{\mathcal T}_0^{\prime} }{{\mathcal T}_0} 
\,  .
\end{equation}
Recalling the low-temperature expression of the Seebeck coefficient associated with a scatterer characterized by a transmission coefficient $\mathcal{T}(\varepsilon)$ \cite{zotti2014}
\begin{equation}
S= - \frac{\pi^2k_{\rm B}^2 T}{3e} \ \frac{{\mathcal T}^{\prime}(\EF)}{{\mathcal T}(\EF)} \, ,
\end{equation}
we see that
\begin{equation}
d_{\rm T} = - \frac{e}{\mu_0} \ T\ S 
\,  .
\end{equation}
The constant $\tilde{\kappa}$ is defined by
\begin{equation}
\label{eq:kappat}
\tilde{\kappa} = \frac{\mathcal{\tilde T}_0}{1-\mathcal{\tilde T}_0} \
\frac{\epsilon L}{\omega \ell} \,  ,
\end{equation}
while $\kappa$ stands for its zero-temperature value.  We introduced the system-dependent parameter
\begin{equation}
\label{eq:omega}
\omega = \epsilon k_{\rm s} L + 2 \epsilon_{\rm B} \,  ,
\end{equation}
where $k_{\rm s}$ is the inverse screening length given in Eq.~\eqref{eq:isl},  while $\epsilon$ and $\epsilon_{\rm B}$ are,  respectively,  the permittivity of the wire and the scatterer.

The linear current-voltage characteristics \eqref{eq:j4plT} is valid up to second-order terms in $e V$.  The absence of quadratic terms is a direct consequence of symmetry requirements.  The linear relationship between the current and the voltage drop in the region of the scatterer implies that any of these two physical quantities can be used in order to fix the strength of the excitation. 

The zero-temperature current-density agrees with the 4-point Landauer formula up to a multiplicative factor (second parenthesis in the right-hand-side).  The finite-temperature correction (third parenthesis) tends to reduce (increase) the linear conductance for a monotonic increasing (decreasing) function $\mathcal{T}(\varepsilon)$ at $\varepsilon=\mu_0$.  Eq.~\eqref{eq:j4plT} is presented in a way to highlight its link with the 4-point Landauer formula,  and as a systematic low-temperature expansion,  only the lowest correction in $\left(k_{\rm B}T\right)^2$ is meaningful. 

The electrostatic potential $\phi_{\rm L,R}(z)$,  the non-equilibrium distribution functions $g_{\rm L,R}(z,\pm v_{\rm L,R})$,  as well as the carrier density $n_{\rm L,R}(z)$,  in the WELT regime,  can be obtained from the results of Appendix \ref{sec:Appltlv},  allowing the determination of the dissipated power considered in Sec.~\ref{subsec:pdcmfp}.  According to Eqs.~\eqref{eq:elpot},  $\phi_{\rm L,R}(z)$ decay away of the scatterer with terms depending on two length scales: $k_{\rm s}$ and $\ell$,  in a similar fashion as $\delta n_{\rm L,R}(z)$.  The detailed features of this decay are not visible in the simplified picture presented in Fig.~\ref{fig:setup}.

\vspace*{1ex}

\subsection{Velocity-independent relaxation-time $ \tau$ in the weak-excitation,  low-temperature regime}
\label{subsec:crt}

The case of a velocity- (or energy-) independent  relaxation-rate 
$\tau$ (where $l(v_{\rm L,R})= v_{\rm L,R} \tau$) does not allow for an exact solution of Eq.~\eqref{eq:jsc},  unlike the velocity-independent mean-free path model treated in the previous chapter.  However,  Eq.~\eqref{eq:jsc} (and the equivalent one for $z>z_{+}$) can be simplified in the WELT regime with a smooth energy-dependence of $\mathcal{T}(\varepsilon)$.  Once the condition that $j$ is $z$-independent is implemented,  a Sommerfeld-like expansion up to order $(k_{\rm B}T)^2$ yields
\begin{widetext}
\begin{align}
\label{eq:jsccrr}
& \int_{-\infty}^{z} {\rm d}z^{\prime} \
\dmul^{\prime}(z') \ 
\exp{\left(\frac{z^{\prime}-z}{l_{0}} \right)} -
 \int_{z}^{z_{-}} {\rm d}z^{\prime}  \
\dmul^{\prime}(z') \
\exp{\left(\frac{z-z^{\prime}}{l_{0}} \right)}
+
\exp{\left(\frac{z-z_{-}}{l_{0}} \right)}
\left\{ 
{\cal A}_{\rm L}(\dmul^{\prime})
- \frac{j}{\zj}
\right\} 
\nonumber \\
& 
\hspace*{0.3cm}
= 
- 
\tilde{\mathcal T}_0 
 \
\frac{eV}{2 \mu_0}
\left\{ 
\frac{eV}{2} - \nu_{\rm R} -
{\cal A}_{\rm R}(\dmur^{\prime}) 
\right\} \
\exp{\left(\frac{z-z_{-}}{l_{0}} \right)} \
\left[\frac{z-z_{-}}{l_{0}} + 1 \right]
- 
\frac{\pi^2}{6} \frac{\left(k_{\rm B}T\right)^2}{\mu_0}
\ l_{0} \
{\mathcal C}_{\rm L}^{\prime}(\dmul^{\prime},\dmur^{\prime};z) 
 \,  .
 \end{align}
 \end{widetext}
As before,  we employed the notation of assigning the sub-index 0 for the value of the energy-dependent functions evaluated at $\mu_0$ ({\it i.e. } $l_0=v_0 \tau$),  and defined the tilded variables as in Eq.~\eqref{eq:tautilde}.  A similar expression (not shown) is obtained for $z>z_{+}$.  The form of the functionals ${\mathcal A}_{\rm L,R}$ and ${\mathcal C}_{\rm L}^{\prime}$ is presented,  respectively,  in  Eqs.~\eqref{eq:functionalAs} and \eqref{eq:functionalCder} of Appendix \ref{sec:AppWELTCRL}.  

We notice that the left-hand-side of Eq.~\eqref{eq:jsccrr} coincides with that of Eq.~\eqref{eq:jsccmfp} if we take $\ell=l_{0}$.
At zero temperature,  the second term in the right-hand-side of Eq.~\eqref{eq:jsccrr} vanishes,  while the first term  (for which $\tilde{\mathcal T}_0$ can be traded by the Fermi energy value
$\mathcal{T}(\varepsilon_{\rm F})$),    is at least of second order in the excitation ({\it i.e.} in $V$). Thus,  the differences in $\delta \mu_{\rm L,R}$ resulting from the two models of velocity-independent mean-free-path and velocity-independent relaxation-time would only differ at $T=0$ by a second-order term in the excitation. For simplicity,  we are not giving the temperature correction of the quadratic contribution in the excitation,  and therefore in the regime  $eV \ll k_{\rm B}T \ll \mu_0$ only the contribution linear in the excitation is meaningful.  For such a regime,  the temperature corrections of terms of higher order in the excitation can be obtained by continuing the order-by-order iteration of \eqref{eq:jsccrr}.  Within the above-specified assumptions,  
the solution for Eq.~\eqref{eq:jsccrr},  and the analogous equation for $\dmur^{\prime}(z)$,  have the form
\begin{widetext}
\begin{equation}
\label{eq:chpotccrr}
\delta \mu_{\rm L,R}^{\prime}(z) = 
\frac{j}{2 {\mathcal J}_0 l_{0}}
\left\{ 1 +
\frac{\pi^2}{24}  
\left(
\frac{k_{\rm B}T}{\mu_0}\right)^2 -
\exp{\left(\pm \ \frac{z-z_{\mp}}{l_{0}} \right)}
\
\left[ 
{\hat d}_{\rm T}
\mp \frac{eV}{2 \mu_0}
\right]
 \
\right\}
\,  ,
\end{equation}
\end{widetext}
where we have defined the thermal factor
\begin{equation}
\label{eq:therfacwh}
{\hat d}_{\rm T} = 
\frac{\pi^2}{6}  \frac{\left(k_{\rm B}T\right)^2}{\mu_0} 
\left(
\frac{2{\mathcal T}_0^{\prime} }{{\mathcal T}_0} -
\frac{1-{\mathcal T}_0 }{\mu_0}
\right)
\,  .
\end{equation}
We label the thermal factor with a hat in order to stress that it is a model-dependent definition,  thus differing from Eq.~\eqref{eq:therfac}. Nevertheless,  except in the transmission plateaus,  we expect 
${\mathcal T}^{\prime}_0 \gg \mathcal{T}_0/\mu_0$,  and therefore the  two definitions coincide.

The first term in the curly bracket of the right-hand-side of \eqref{eq:chpotccrr} yields the dominant contribution to $\delta \mu_{\rm L,R}^{\prime}(z)$,  which coincides with the result of Eq.~\eqref{eq:chpotcmfp} upon the replacement of $\ell$ by $l_0$.
Comparing \eqref{eq:chpotcmfp} and \eqref{eq:chpotccrr},  we see that going from the model with a velocity-independent mean-free-path to that with a velocity-independent relaxation-time results in two corrections in the functional form of the electro-chemical potential.  One the one hand,  there appears a space-independent temperature correction (second term within the curly bracket), representing the Sommerfeld-like correction to the Drude conductivity of the wire.  Such a correction results,  at finite temperature,  in an increase of the slope of $-\mu_{\rm L,R}(z)$ with respect to the case presented in Fig. ~\ref{fig:setup} for a velocity-independent mean-free-path.  On the other hand,  the linear drop of the electro-chemical potential in the wires is perturbed around the scatterer (third term within the curly bracket) with temperature-dependent and quadratic contributions that break the left-right symmetry.   

The form \eqref{eq:chpotccrr} of the electro-chemical potential allows to write the space-independent factors $g_{\rm L,R}^{(1)}(\mp v_{\rm L,R})$ defined in Eq.~\eqref{eq:g1Landg1R} as 
\begin{widetext}
\begin{align}
\label{eq:g1LRcT}
g_{\rm L,R}^{(1)}(\mp v_{\rm L,R})  = &
\pm f^{\prime}(\varepsilon_{\rm L,R}^{\rm c}) \ \frac{j}{2 \zj}
\left\{ 
\left[
1 +
\frac{\pi^2}{24} 
\left(
\frac{k_{\rm B}T}{\mu_0}\right)^2 
\right]   \frac{v_{\rm L,R}}{v_0} -
\left[ 
{\hat d}_{\rm T}
\mp \frac{eV}{2 \mu_0}
\right]
 \frac{v_{L,R}}{v_{0}+v_{L,R}}
\right\}
\nonumber
\\
&
\mp \mathcal{T}(\varepsilon)
\Biggl(
f(\ecL) - f(\ecR) - \nu_{\rm L} 
\ f^{\prime}(\ecL) +
\nu_{\rm R} \ f^{\prime}(\ecR) 
\nonumber
\\
&
+
f^{\prime}(\varepsilon_{\rm L}^{\rm c}) \ \frac{j}{2 \zj}
\left\{ 
\left[
1 +
\frac{\pi^2}{24} 
\left(
\frac{k_{\rm B}T}{\mu_0}\right)^2 
\right]  \ \frac{v_{\rm L}}{v_0} \ -
\
\left[ {\hat d}_{\rm T}
- \frac{eV}{2 \mu_0}
\right]
\ \frac{v_{L}}{v_{0}+v_{L}}
\right\}
\nonumber
\\
&
+
f^{\prime}(\varepsilon_{\rm R}^{\rm c}) \ \frac{j}{2 \zj}
\left\{ 
\left[
1 +
\frac{\pi^2}{24} 
\left(
\frac{k_{\rm B}T}{\mu_0}\right)^2 
\right]  \ \frac{v_{\rm R}}{v_0} \ -
\
\left[ {\hat d}_{\rm T}
+ \frac{eV}{2 \mu_0}
\right]
\ \frac{v_{R}}{v_{0}+v_{R}}
\right\}
\Biggr)
 \,  .
\end{align}
\end{widetext}

The particle-conserving condition \eqref{eq:pccgen} is now a simple statement that the non-equlibrium carrier density $n_{\rm L,R}(z)$ coincides with its local equilibrium counterpart $n^{(\rm le)}_{\rm L,R}(z)$,  defined in \eqref{eq:nledef}.  Therefore,  according to Eqs.~\eqref{eq:nle} and \eqref{eq:deltanle},  the form of Poisson's equation is the same as in \eqref{eq:Poisson},  up to the identifications presented in Eq.~\eqref{eq:identifications}.  The electrostatic problem for a velocity-independent relaxation-time is then analogous as in the case of a velocity-independent mean-free-path treated in Appendix \ref{sec:appendixSCTEP}.  The resulting corrections between the two models are obtained in Appendix \ref{sec:AppWELTCRL},  leading to the current-voltage characteristics
\begin{widetext}
\begin{equation}
\label{eq:j4plTtau}
j =  \frac{G_0}{A} \ V
\left( \frac{\mathcal{\tilde T}_0}{1-\mathcal{\tilde T}_0} \right) 
\left(
\frac{1+2\epsilon_{\rm B}/\omega}{1+\tilde{\kappa} } 
\right)
\left(1 + {\hat d}_{\rm T} \ \frac{\kappa}{1+\kappa} \  \frac{k_{\rm s}l_0}{k_{\rm s}l_0+1}
+
 \frac{\pi^2}{12} \left(\frac{k_{\rm B}T}{ \mu_0}\right)^2  \
\frac{ 1}{1+\kappa} 
 \left(
  {\mathcal T}_0 -
    \frac{\kappa}{2} 
 \right) 
\right)
\,  ,
\end{equation}
\end{widetext}
where ${\tilde \kappa}$ is defined by Eq.~\eqref{eq:kappat} but now using $l_0$ instead of $\ell$.  Comparing the current-voltage characteristics of the two models,  {\it i.e.}  Eqs.~\eqref{eq:j4plT} and \eqref{eq:j4plTtau},  we see that they coincide at zero temperature (when taking $\ell=l_0$),  and thus only differ in their low-temperature correction.

\section{Dissipated power}
\label{sec:dp}

The Boltzmann equation allows to express the power density that the carriers dissipate into the environement (excitation of phonons in the host material,  eventually leading to heating) as \cite{rokni1995}
\begin{equation}
\label{eq:ddpRL}
p_{\rm L,R}^{\rm q}(z)= p_{\rm L,R}^{\rm w}(z) + p_{\rm L,R}^{\rm u}(z) \,  ,
\end{equation}
with
\begin{widetext}
\begin{subequations}
\label{eq:dpdcomponents}
\begin{align}
\label{eq:dpdfcomponent}
p_{\rm L,R}^{\rm w}(z) & = 
 \frac{1}{\pi A \hbar} \ e \phi^{\prime}(z) \int_{0}^{\infty} {\rm d}
v_{\rm L,R} \ \varepsilon_{\rm L,R}^{\rm c} \
\frac{{\partial} }{{\partial} v_{\rm L,R}} 
\biggl(  
g_{\rm L,R}(z,v_{\rm L,R})-
g_{\rm L,R}(z,-v_{\rm L,R})
\biggr)   = -  j \ \phi^{\prime}_{\rm L,R}(z)
 \,  ,
\\
\label{eq:dpdscomponent}
p_{\rm L,R}^{\rm u}(z) & = 
- \frac{m}{\pi A \hbar} \int_{0}^{\infty} {\rm d}v_{\rm L,R} \ \varepsilon_{\rm L,R}^{\rm c} \ v_{\rm L,R}  \
\frac{{\partial} }{{\partial} z} 
\biggl(  
g_{\rm L,R}(z,v_{\rm L,R})-
g_{\rm L,R}(z,-v_{\rm L,R})
\biggr)   
 \,  .
\end{align}
\end{subequations}
\end{widetext}
The second equality of Eq.~\eqref{eq:dpdfcomponent} follows from the form \eqref{eq:jdefa} of the current density.  The above decomposition can be viewed as a statement of the first principle of thermodynamics applied to an Eulerian description of an infinitesimal sector of the wire for the energy flow accompanying the moving carriers.  The power that electrons dissipate into the environment $p_{\rm L,R}^{\rm q}(z)$,  which translates into (outgoing) heat,  equals the (incoming) work done by the electrostatic field on the carriers (first contribution,  $p_{\rm L,R}^{\rm w}(z)$) plus the opposite of the variation of the internal energy,  {\it i.e.}~the energy-flow across the sector (second contribution,  $p_{\rm L,R}^{\rm u}(z)$).  The presence of the scatterer results in a non-homogeneity of the energy-flow,  giving rise to $p_{\rm L,R}^{\rm u}(z)$,  and also modifies the electrostatic field,  thus driving $p_{\rm L,R}^{\rm w}(z)$ away from its Drude value $p^{\rm D} = -  j \ \phi^{\prime}_{\rm L,R}(\mp \infty)$
(which,  at zero temperature or for an energy-independent mean-free-path is given by $- j^2/(2e\zj l)$).  Leaving aside this trivial Drude dissipation power,  we are interested in the excess power dissipation due to the presence of the scatterer,  that we write as
\begin{equation}
\label{eq:ddpRLs}
p_{\rm L,R}^{\rm s}(z)= p_{\rm L,R}^{\rm q}(z) - p^{\rm D} =
p_{\rm L,R}^{\rm f}(z) + p_{\rm L,R}^{\rm u}(z) \,  ,
\end{equation}
where we have defined the field component $p_{\rm L,R}^{\rm f}(z)= p_{\rm L,R}^{\rm w}(z) - p^{\rm D} $.  

The identity \eqref{eq:useful}, together with the particle-conserving condition \eqref{eq:pccgen},  allow to trade $\varepsilon_{\rm L,R}^{\rm c}$ by 
$\varepsilon_{\rm L,R}^{\rm c}+c_{\rm L,R}$ in 
Eq.~\eqref{eq:dpdscomponent},  where $c_{\rm L,R}$ are arbitrary energy-independent constants.   Using this freedom,  and the form \eqref{eq:distributions} of the distribution function,  we can write
\begin{widetext}
\begin{subequations}
\label{eq:pe}
\begin{align}
\label{eq:pe1}
p_{\rm L}^{\rm u}(z) & = 
\frac{1}{\pi A \hbar} \int_{0}^{\infty} 
{\rm d}\ecL \
\left[
\ecL+c_{\rm L}
\right] \
\frac{1}{l(v_{\rm L})} \
\Biggl[ 
f^{\prime}(\ecL) 
\Biggl\{
\int_{-\infty}^{z} {\rm d}z^{\prime} \
\dmul^{\prime}(z') \
\exp{\left(\frac{z^{\prime}-z}{l(v_{\rm L})} \right)} 
\nonumber \\
& \hspace{1.5cm}
-
\int_{z}^{z_{-}} {\rm d}z^{\prime} \
\dmul^{\prime}(z') \
\exp{\left(\frac{z-z^{\prime}}{l(v_{\rm L})} \right)} \
\Biggr\}
+
g_{\rm L}^{(1)}(-v_{\rm L}) \
\exp{\left(\frac{z-z_{-}}{l(v_{\rm L})} \right)}
\Biggr]
 \,  ,
\\
\label{eq:pe2}
p_{\rm R}^{\rm u}(z) & = 
\frac{1}{\pi A \hbar} \int_{0}^{\infty} 
{\rm d}\ecR \
\left[
\ecR+c_{\rm R}
\right] \
\frac{1}{l(v_{\rm R})} \
\Biggl[ 
f^{\prime}(\ecR) 
\Biggl\{
\int_{z_{+}}^{z} {\rm d}z^{\prime} \
\dmur^{\prime}(z') \
\exp{\left(\frac{z^{\prime}-z}{l(v_{\rm R})} \right)} \
\nonumber \\
& \hspace{1.5cm}
-
\int_{z}^{\infty} {\rm d}z^{\prime} \
\dmur^{\prime}(z') \
\exp{\left(\frac{z-z^{\prime}}{l(v_{\rm R})} \right)} \
\Biggr\}
+
g_{\rm R}^{(1)}(v_{\rm R}) \
\exp{\left(\frac{z_{+}-z}{l(v_{\rm L})} \right)}
\Biggr]
 \,  .
\end{align}
\end{subequations}
\end{widetext}
In our discussion of Eq.~\eqref{eq:j4p} of Sec.~\ref{subsec:cmfp},  we introduced the auxiliary chemical potentials $\mu_{\rm L,R} = \mu_0 \pm e V/2$ in order to make the connection with the results obtained within the scattering approach.  Since we are interested in discussing such a connection for the dissipation power \cite{SciPostPhys2022},  we will chose in the sequel $c_{\rm L,R}= - \mu_0$,  and then 
$\varepsilon_{\rm L,R}^{\rm c}+c_{\rm L,R} = \varepsilon - \mu_{\rm L,R}$.

Two salient features of power dissipation on nano-structures have been discussed in the literature: the asymmetry between the dissipated power at both sides of the scatterer \cite{lee2013,SciPostPhys2022,zotti2014,sanchez2015,NicoPRB},  and
the existence of hot regions away from the scatterer where the local temperature is increased with respect to the background one
\cite{zeldov_pc2019,Mirlin2019,SciPostPhys2022,NicoPRB}.  The approximate solutions obtained for the two models discussed in the previous chapters allow to address both of these physical effects.  While for the dissipation asymmetry we have to consider spatial integrals of $p_{\rm L,R}^{\rm s}(z)$,  the heating (or cooling) spots are determined by the maxima (or minima) of the dissipated power density $p_{\rm L,R}^{\rm q}(z)$ (which are obviously the same than for the excess power dissipation $p_{\rm L,R}^{\rm s}(z)$ due to the presence of the scatterer). 

We define,  respectively,  the total excess power dissipation,  and its asymmetry between the right and left electrodes,  by
\begin{equation}
\mathcal{P}_{\rm T,A}^{\rm s} = \mathcal{P}_{\rm R}^{\rm s} \pm \mathcal{P}_{\rm L}^{\rm s}  \,  ,
\end{equation}
with
\begin{subequations}
\label{eq:ptotal}
\begin{align}
\label{eq:ptotalL}
\mathcal{P}_{\rm L}^{\rm s} & =  \int_{-\infty}^{z_{-}} 
{\rm d}z \ p_{\rm L}^{\rm s}(z) \,  ,
\\
\label{eq:ptotalR}
\mathcal{P}_{\rm R}^{\rm s} & =  \int_{z_{+}}^{\infty} 
{\rm d}z \ p_{\rm R}^{\rm s}(z)
 \,  .
\end{align}
\end{subequations}
Analogous definitions of $\mathcal{P}_{\rm T,A}^{\rm f}$,  
$\mathcal{P}_{\rm T,A}^{\rm u}$,  $\mathcal{P}_{\rm L,R}^{\rm f}$,  and
$\mathcal{P}_{\rm L,R}^{\rm u}$ are implemented by using the components $p_{\rm L,R}^{\rm f}(z)$ and $p_{\rm L,R}^{\rm u}(z)$ instead of $p_{\rm L,R}^{\rm s}(z)$ in Eqs.~\eqref{eq:ptotal}.

\vspace*{1cm}

\subsection{Power dissipation with a velocity-independent mean-free-path $\ell$}
\label{subsec:pdcmfp}

In the case of a velocity-independent mean-free-path $\ell$,  Eqs.~\eqref{eq:pe}  become
\begin{equation}
\label{eq:ldp}
p_{\rm L,R}^{\rm u}(z)  = \frac{\mathcal{P}_{\rm L,R}^{\rm u}}{ A \ell} \ 
 \exp{\left(\pm \ \frac{z-z_{\mp}}{\ell} \right)} 
 \,  ,
\end{equation}
with 
\begin{widetext}
\begin{equation}
\label{eq:dpLR2}
\mathcal{P}_{\rm L,R}^{\rm u}  =  \frac{m}{\pi \hbar} \ 
\int_{0}^{\infty}  {\rm d}v_{\rm L,R} 
\left[ \varepsilon_{\rm L,R}^{\rm c} - \mu_{0} \right]  v_{\rm L,R}  
\left[ g_{\rm L,R}^{(1)}(\mp v_{\rm L,R}) \pm \frac{j}{2 \zj} f^{\prime}(\varepsilon_{\rm L,R}^{\rm c}) \right] \,  .
\end{equation}
\end{widetext}
It is straightforward to verify that 
\begin{equation}
\mathcal{P}_{\rm T}^{\rm u}  = A \ j  \ V \, , 
\end{equation}
indicating that the potential drop $V$ induced by the scatterer is associated with the component $\mathcal{P}_{\rm T}^{\rm u}$ of the total dissipated power according to the standard Ohm-Joule formula.  However, it is worth to remark that the potential drop $V$ occurs within a distance $\ell$ of the scatterer,  while the drop of the electrostatic potential across the scatterer $\Delta \phi$ is a fraction of $V$ (see Eq. \eqref{eq:potdroplT}),  and that the component $\mathcal{P}_{\rm T}^{\rm f}$ is also originated by the presence of the scatterer.  

The form \eqref{eq:g1LRsim} of the space-independent factors $g_{\rm L,R}^{(1)}(\mp v_{\rm L,R})$ leads to  
%
%
%
\begin{widetext}
\begin{align}
\label{eq:apdcomponents}
\mathcal{P}_{\rm L,R}^{\rm u} & = \mp \frac{m}{\pi \hbar} 
\int_{0}^{\infty} {\rm d}v_{\rm L,R}  \ v_{\rm L,R}
\left[ \varepsilon - \mu_{\rm L,R} \right] \mathcal{T}(\varepsilon) 
\left[ f(\ecL) - f(\ecR) \right]
\nonumber
\\
 & \mp \frac{m}{\pi \hbar} \ \frac{j}{2 \zj} 
\int_{0}^{\infty} {\rm d}v_{\rm L,R}  \ v_{\rm L,R} \ 
\left[ \varepsilon - \mu_{\rm L,R} \right]  \mathcal{T}(\varepsilon)
\left[f^{\prime}(\ecL) + f^{\prime}(\ecR) \right]
\nonumber
 \\
  & \pm \frac{m}{\pi \hbar}
\int_{0}^{\infty} {\rm d}v_{\rm L,R}  \ v_{\rm L,R} \ 
\left[ \varepsilon - \mu_{\rm L,R} \right]  \mathcal{T}(\varepsilon)
\left[
f^{\prime}(\ecL) \ \nu_{\rm L} - f^{\prime}(\ecR) \ \nu_{\rm R} \right]
 \,  .
\end{align}
\end{widetext}
We have neglected the contribution to $\mathcal{P}_{\rm L,R}^{\rm u}$ from the terms of  \eqref{eq:dpLR2} without $\mathcal{T}(\varepsilon)$,  since in the corresponding integrals we have the product of an even and and odd function of $\varepsilon-\mu_0$. 
The first contribution to $\mathcal{P}_{\rm L,R}^{\rm u}$ corresponds to the power dissipation obtained within the scattering approach between left  and right reservoirs with chemical potentials $\mu_{\rm L,R}$,  as considered in Refs.~\cite{SciPostPhys2022,zotti2014}.   The two remaining terms represent the additional dissipation arising from the inelastic scattering in the wires,  and they will be evaluated in the WELT regime in Sec.~\ref{subsec:sadpltl},  together with their resulting contribution to the power dissipation asymmetry.  

The spatial asymmetry associated with the field-induced dissipated power $\mathcal{P}_{\rm A}^{\rm f}$ follows from the general form 
\eqref{eq:elpot} of $\phi_{\rm L}(z)$ and \eqref{eq:ALAR} of the parameters $A_{\rm L,R}$,  and can be expressed as
\begin{equation}
\label{eq:fdpa}
\mathcal{P}_{\rm A}^{\rm f}  = - j A \ \frac{1}{2e|\mathcal{F}_{0}|} \  \frac{k_{\rm s} \ell}{k_{\rm s} \ell+1} \left( q_{\rm R} - q_{\rm L} \right) 
 \,  .
\end{equation} 
In Sec.~\ref{subsec:sadpltl} we will analyze this component of the power dissipation asymmetry for the WELT regime,  comparing it with the contribution $\mathcal{P}_{\rm A}^{\rm u}$.   

From Eqs.~\eqref{eq:ddpRL},  \eqref{eq:dpdcomponents}, \eqref{eq:ldp},  and \eqref{eq:dpLR2},  we can write 
\begin{equation}
p_{\rm L,R}^{\rm q}(z) = -  j \ \phi^{\prime}_{\rm L,R}(z)+
 \frac{\mathcal{P}_{\rm L,R}^{\rm u} }{A \ell} \
 \exp{\left(\pm \ \frac{z-z_{\mp}}{\ell} \right)} 
 \,  .
\end{equation}
Using the form \eqref{eq:elpot} of the electrostatic potential,  the possible appearance of heating (or cooling) spots is determined by the following vanishing conditions  
\begin{widetext}
\begin{equation}
\label{eq:condhs}
\frac{{\rm d} p_{\rm L,R}^{\rm s}(z)}{{\rm d} z} = -  j k_{\rm s}^2 A_{\rm L,R}  \exp{\left(\pm k_{\rm s}\left[ z-z_{\mp} \right] \right)} \pm
\left\{
 \frac{\mathcal{P}_{\rm L,R}^{\rm u} }{A \ell^2} - 
 \frac{4\pi e}{\epsilon} \  \frac{j\ q_{\rm L,R}}{k_{\rm s}^2 \ell^2-1}
 \right\}
 \exp{\left(\pm \ \frac{z-z_{\mp}}{\ell} \right)} = 0
 \,  ,
\end{equation}
leading to the solutions
\begin{equation}
\label{eq:extreme}
{\bar z}_{\rm L,R} = z_{\mp} \mp 
\left(
\frac{\ell}{k_{\rm s} \ell -1} 
\right) \
\ln {\left[ \frac{\pm \ A_{\rm L,R} \  k_{\rm s}^2 \ell^2}
{ \frac{\mathcal{P}_{\rm L,R}^{\rm u} }{ j A}  -
\frac{1}{2e| \mathcal{F}_0|} \
 \frac{k_{\rm s}^2 \ell^2}{k_{\rm s}^2 \ell^2-1} \ q_{\rm L,R}  }\right]} \,  ,
\end{equation}
\end{widetext}
at the left (right) of the scatterer,  provided that
\begin{equation}
\label{eq:cond1}
\pm \frac{1}{A_{\rm L,R}}
\left(
 \frac{\mathcal{P}_{\rm L, R}^{\rm u} }{ j A}  -
\frac{1}{2e| \mathcal{F}_0|} \
 \frac{k_{\rm s}^2 \ell^2}{k_{\rm s}^2 \ell^2-1} \ q_{\rm L,R} 
\right) > 0 \,  , 
\end{equation}
and
\begin{equation}
\label{eq:cond2}
A_{\rm L,R} \ k_{\rm s}^2 \ell^2 > \pm
\left(
 \frac{\mathcal{P}_{\rm L,R}^{\rm u} }{ j A}  -
\frac{1}{2e| \mathcal{F}_0|} \
 \frac{k_{\rm s}^2 \ell^2}{k_{\rm s}^2 \ell^2-1} \ q_{\rm L,R}
 \right) 
 \,  .
\end{equation}
The fulfillment of the previous conditions dictates the presence of a thermal spot at the left (right) of the scatterer.  Its nature,  {\em i.e} heating or cooling,  depends on the sign of
\begin{widetext}
\begin{equation}
\frac{{\rm d}^2 p_{\rm L,R}^{\rm s}(z)}{{\rm d} z^2} {\Big |}_{z={\bar z}_{\rm L,R}} 
 = j \
\left(\frac{k_{\rm s} \ell-1}{\ell}
\right) \
\left(
\frac{1}{2e| \mathcal{F}_0|} \
 \frac{k_{\rm s}^2 \ell^2}{k_{\rm s}^2 \ell^2-1} \ q_{\rm L,R} -
  \frac{\mathcal{P}_{\rm L,R}^{\rm u} }{ j A}  
\right)
 \exp{\left(\pm \ \frac{{\bar z}_{\rm L,R}-z_{\mp}}{\ell} \right)} 
 \,  , 
\end{equation}
 \end{widetext}
which is itself given by that of the second parenthesis in the right-hand-side of the above expression.  Given the conditions \eqref{eq:cond1},  such a sign can be determined from that of $A_{\rm L,R}$.

The solutions \eqref{eq:extreme} for the thermal spots ${\bar z}_{\rm L,R}$ and the conditions \eqref{eq:cond1}-\eqref{eq:cond2} for their existence can,  in principle,  be discussed in the generic case by using the expression \eqref{eq:j4pex} of the current density,  together with the resulting forms of $q_{\rm L,R}$.  However,  in order to simplify such a discussion,  we would restrict ourselves in the sequel to the limited scenario of the WELT regime treated in the next chapter.

\subsubsection{Spatial asymmetry and thermal spots in the weak-excitation,  low-temperature regime}
\label{subsec:sadpltl}

Working in the WELT regime allows to simplify the different terms of 
Eq.~\eqref{eq:apdcomponents},  leading to a dissipated power-density component arising from the energy-flow inhomogenity given by
\begin{widetext}
\begin{equation}
\label{eq:pLRecmfp}
 \mathcal{P}_{\rm L,R}^{\rm u}  = \frac{1}{2} \ A \ j  \ V 
 \mp \frac{\pi^2}{3} \left(k_{\rm B}T \right)^2
\frac{\mathcal{T}^{\prime}_0}{\mathcal{T}_0} \ \frac{Aj}{e} \mp
\frac{A j eV^2}{8 \mu_0} \ \mathcal{T}_0 \
\frac{k_{\rm s} \ell}{k_{\rm s} \ell+1} \mp 
\left( e V \right)^2
 \left\{
 \frac{A j}{4 e \mathcal{T}_0} -   \frac{e V}{6\pi \hbar}
\right\}   \mathcal{T}^{\prime}_0
 \,  ,
\end{equation}
and the corresponding asymmetry
\begin{equation}
\label{eq:paecmfp}
\mathcal{P}_{\rm A}^{\rm u} =
\frac{2 \pi^2}{3} \left(k_{\rm B}T \right)^2
\frac{\mathcal{T}^{\prime}_0}{\mathcal{T}_0} \ \frac{j}{e} +
\frac{A j eV^2}{4 \mu_0} \ \mathcal{T}_0 \
\frac{k_{\rm s} \ell}{k_{\rm s} \ell+1} + 
\left( e V \right)^2
 \left\{
 \frac{A j}{2 e \mathcal{T}_0} -   \frac{e V}{3\pi \hbar}
\right\}   \mathcal{T}^{\prime}_0
  \,  .
\end{equation}
\end{widetext}
At zero temperature,  $\mathcal{P}_{\rm L,R}^{\rm u} $ is quadratic with the excitation,  and the dominant term is left-right symmetric,  while the asymmetry $\mathcal{P}_{\rm A}^{\rm u}$ scales with the third power of the excitation.  At finite temperature,  both quantities present a contribution that is of first order in the excitation.  

The contribution to $\mathcal{P}_{\rm A}^{\rm u}$ arising from the first term in the right-hand-side of 
Eq.~\eqref{eq:apdcomponents} has been shown \cite{SciPostPhys2022,zotti2014} to be
$
(\mathcal{T}^{\prime}_0 e V/6 \pi \hbar)
\{
\left(e V\right)^2 + 
4 \pi^2 \left(k_{\rm B}T \right)^2  
\} 
$,
which is positive at zero temperature and grows with a finite temperature when the function $\mathcal{T}(\varepsilon)$ is monotonically increasing with energy.  This asymmetry rendering more power dissipation downstream of the scatterer is reinforced by the remaining terms in Eq.~\eqref{eq:apdcomponents},  as can be easily checked from \eqref{eq:paecmfp}.  However,  the field-induced contribution to the power dissipation asymmetry $\mathcal{P}_{\rm A}^{\rm f}$ \eqref{eq:fdpa},  when using the results of Appendix \ref{sec:Appltlv},  results in  
\begin{equation}
\label{eq:paf}
\mathcal{P}_{\rm A}^{\rm f} =
-  \
\frac{(A j)^2 V}{e} 
\ \frac{\pi \hbar}{4 \mu_0} \
\frac{k_{\rm s} \ell}{k_{\rm s} \ell+1} 
 \,  ,
\end{equation}
which does not depend on the monotonicity of $\mathcal{T}(\varepsilon)$ and it is negative,  thus reducing the previously discussed effect of more power dissipation downstream of the scatterer.  
We see that, including the dissipation in the vicinity of the scatterer,  the positive power asymmetry predicted from the scattering approach  for $\mathcal{T}^{\prime}(\EF) > 0$ at low temperatures  \cite{SciPostPhys2022},  might increase or decrease,  according to the various physical parameters characterizing the scatterer and the connecting wires.  

The discussion of the existence and location of thermal spots is considerably simplified in the WELT regime,  where we use the form \eqref{eq:qLRlT} and \eqref{eq:ALARlT} of the parameters $q_{\rm L,R}$ and $A_{\rm L,R}$,  respectively.  Moreover,  in the metallic case we expect the screening length to be much smaller than the mean-free-path,  {\it i.e. } $k_{\rm s} \ell \gg 1$,  and for a ballistic scatterer we have $\ell \gg L$.  Within the previous hypothesis,  the conditions \eqref{eq:cond1} for the existence of a thermal spot at the left (right) of the scatterer can be expressed as

\begin{equation}
\label{eq:condcsrw}
\mp j \ \frac{\pi A \hbar}{4 e \mu_0} +
\frac{1}{2} \ \frac{{\mathcal T}_0}{1-\mathcal{T}_0 } \ d_{\rm T} \
\frac{1+2\epsilon_{\rm B}/\omega}{1+\kappa} > 1
\,  ,
\end{equation}
%
by only keeping the lowest-order terms in excitation and temperature.  While finite-temperature favors the emergence of a thermal spot both, at the left and at the right of the scattarer,  the excitation strength acts in an asymmetric way,  favoring such an emergence downstream of the scatterer and hindering it upstream.  The role of temperature is crucial,  since at zero temperature there is no thermal spot at the left of the scaterer,  and the critical current to have a thermal spot at the right of the scatterer is too high to be realistic.  

Under the above assumptions,  the condition  \eqref{eq:cond2} is fulfilled for a large range of parameters and a cooling spot is located at
\begin{widetext}
\begin{equation}
\label{eq:hotspotlT}
{\bar z}_{\rm L,R} = z_{\mp} \mp k_{\rm s}^{-1} \ 
\ln {\left[ \frac{k_{\rm s}^2 \ell^2
\left(
\frac{2\epsilon_{\rm B}/\omega-\kappa}{1+\kappa }  + 
\frac{1}{2} \ \frac{{\mathcal T}_0}{1-\mathcal{T}_0 } \ d_{\rm T} \
\frac{1+2\epsilon_{\rm B}/\omega}{1+\kappa}
\left[
1   -
\frac{\epsilon k_{\rm s} L}{\omega(1+\kappa)}
\right]
 \right)
}
{\mp j \frac{\pi A \hbar}{4e\mu_0} 
+ \frac{1}{2} \ \frac{{\mathcal T}_0}{1-\mathcal{T}_0 } \ d_{\rm T} \
\frac{1+2\epsilon_{\rm B}/\omega}{1+\kappa}
 -1}\right]} \,  .
\end{equation}

The length-scale for the separation of the cooling spot and the scatterer is the screening length,  but the logarithmic factor depending on $(k_{\rm s} \ell)^2$ drives the cooling spot away of the scatterer.  Depending on the values of different parameters,  finite temperature could drive the cooling spot further away from $z_{\mp}$ and make the dissipated power density to attain a more pronounced local minimum (with smaller values). 


\subsection{Power dissipation with a velocity-independent relaxation-time $ \tau$ in the weak-excitation,  low-temperature regime}
\label{subsec:pdcrt}

In the case of a velocity-independent relaxation-time $\tau$,  within the WELT regime,  a Sommerfeld-like integration of Eqs.~\eqref{eq:pe}  leads to a dissipated power density component arising from the energy-flow inhomogenity given by
%
\begin{equation}
\label{eq:ldpcrt}
p_{\rm L,R}^{\rm u}(z)  = 
\frac{j}{e l_0} \ 
 \exp{\left(\pm \ \frac{z-z_{\mp}}{l_0} \right)} 
\left\{   \frac{eV}{2} \mp 
{\hat d}_{\rm T}
 +
 \frac{\pi^2}{12} \left( \frac{k_{\rm B}T}{\mu_0} \right)^2
  \left( \frac{z-z_{\mp}}{l_0}
\right)
 \
\right\}
\,  ,
\end{equation}
together with the associated power dissipated in the left (right) wire 
\begin{equation}
\label{eq:dpeicrt}
\mathcal{P}_{\rm L,R}^{\rm u}  =  \frac{1}{2} \ A \ j \ V \mp
\frac{\pi^2}{3} \left(k_{\rm B}T\right)^2
\frac{A j}{e}  
\left(
\frac{{\mathcal T}_0^{\prime} }{{\mathcal T}_0} -
\frac{1}{2\mu_0} 
\left(
\frac{1}{2} - {\mathcal T}_0 
\right)
\right)
\pm
\frac{A j eV^2}{4 \mu_0} \ \mathcal{T}_0 \
\frac{1}{k_{\rm s} l_0+1}
\mp 
\left( e V \right)^2
 \left\{
 \frac{A j}{4 e \mathcal{T}_0} -   \frac{e V}{6\pi \hbar}
\right\}   \mathcal{T}^{\prime}_0
 \,  .
\end{equation}
The resulting asymmetry is
\begin{equation}
\label{eq:adpeicrt}
\mathcal{P}_{\rm A}^{\rm u}  =  
\frac{2\pi^2}{3} \left(k_{\rm B}T\right)^2
\frac{A j}{e}  
\left(
\frac{{\mathcal T}_0^{\prime} }{{\mathcal T}_0} -
\frac{1}{2\mu_0} 
\left(
\frac{1}{2} - {\mathcal T}_0 
\right)
\right)
-
\frac{A j eV^2}{2 \mu_0} \ \mathcal{T}_0 \
\frac{1}{k_{\rm s} l_0+1}
+ 
\left( e V \right)^2
 \left\{
 \frac{A j}{2 e \mathcal{T}_0} -   \frac{e V}{3\pi \hbar}
\right\}   \mathcal{T}^{\prime}_0
 \,  ,
\end{equation}
\end{widetext}
while the field component of the asymmetry is
\begin{equation}
\label{eq:paftau}
\mathcal{P}_{\rm A}^{\rm f} =
-  \
\frac{(A j)^2 V}{e} 
\ \frac{\pi \hbar}{2 \mu_0} \
\frac{k_{\rm s} \ell}{k_{\rm s} \ell+1} 
 \,  .
\end{equation}
Similarly to the case of a velocity-independent mean-free-path discussed after Eq.~\eqref{eq:paf},  the power asymmetry obtained within the scattering approach can be substantially modified by the effect of $\mathcal{P}_{\rm A}^{\rm f}$.  A similar departure from Landauer scattering theory was detected for the different regimes studied in Ref.~\cite{NicoPRB} demonstrating the need to go beyond such an approach in order to describe the consequences of dissipation in the neighborhood of the scatterer. 

While in Eqs.~\eqref{eq:dpeicrt}-\eqref{eq:paftau} we have included terms that are cubic in the excitation (which are the ones that determine the power dissipation asymmetry at zero temperature), such terms have not been reported in Eq.~\eqref{eq:ldpcrt}, since they are not relevant for the determination of the thermal spots.

The possible appearance of thermal spots is determined by the following vanishing conditions  
\begin{widetext}
\begin{align}
\label{eq:condhscrt}
\frac{{\rm d} p_{\rm L,R}^{\rm s}(z)}{{\rm d} z} = &
-  j k_{\rm s}^2 A_{\rm L,R}  \exp{\left(\pm k_{\rm s}\left[ z-z_{\mp} \right] \right)} \
-
\
\frac{j}{e}
\Biggl\{
\frac{1}{l_0^2}
\left[ 
\frac{\pi^2}{6} \left(k_{\rm B}T\right)^2
\left(
\frac{2{\mathcal T}_0^{\prime} }{{\mathcal T}_0} -
\frac{3-2{\mathcal T}_0 }{2\mu_0} \mp
 \frac{1}{2 \mu_0} \left( \frac{z-z_{\mp}}{l_0} \right)
\right)
\mp \frac{eV}{2}
\right]
\nonumber
\\
&
 \pm \
 \frac{j}{e} \
 \frac{\pi A \hbar}{2} \  \frac{k_{\rm s}^2}{k_{\rm s}^2 \l_0^2-1} 
 \left[ 
\frac{\pi^2}{6}  \frac{\left(k_{\rm B}T\right)^2}{\mu_0} 
\left(
\frac{2{\mathcal T}_0^{\prime} }{{\mathcal T}_0} -
\frac{1-{\mathcal T}_0 }{\mu_0}
\right)
\mp \frac{eV}{2 \mu_0}
\right]
\Biggr\}
 \exp{\left(\pm \ \frac{z-z_{\mp}}{\ell} \right)} = 0
 \,  ,
\end{align}
which cannot be explicitly solved.  However,  in the case 
$k_{\rm s} \l_0 \gg 1$,  we can linearize around the solutions 
$z_{\rm L,R}^{{\rm scl}}$ obtained in the strong Coulomb limit in Appendix~\ref{sec:sclcrt},  yielding
\begin{equation}
{\bar z}_{\rm L,R} = z_{\rm L,R}^{{\rm scl}} + \Delta z_{\rm L,R} \, ,
\end{equation}
with $z_{\rm L,R}^{{\rm scl}}$ given by \eqref{eq:hotspotlTcrt} and
\begin{equation}
\label{eq:deltaz}
 \Delta z_{\rm L,R} = l_0 \ e V
\left(\frac{2\epsilon_{\rm B}/\omega-
\kappa}{1+\kappa} 
\right)
\left(
\frac{2{\mathcal T}_0^{\prime} }{1-{\mathcal T}_0} -
\frac{{\mathcal T}_0 }{\mu_0}
\mp \frac{4 eV}{\pi^2 \left(k_{\rm B}T\right)^2} \
\frac{{\mathcal T}_0 }{1-{\mathcal T}_0} 
\right) \, .
\end{equation}
\end{widetext}
The conditions for having a cooling (heating) spot in the left (right) lead established in App.~\ref{sec:sclcrt} should now incorporate the corrections $\Delta z_{\rm L,R}$ in the requirements
${\bar z}_{\rm L} < z_{-}$ and ${\bar z}_{\rm R} > z_{+}$.  Finite temperatures were shown to be crucial for obtaining a physically meaningful $z_{\rm L,R}^{{\rm scl}}$, and this is also the case for the corrections $ \Delta z_{\rm L,R}$.  For the two wires,  the voltage drop $V$ acts as to drive the thermal spot away from the scatterer,  thus favoring its appearance.  For a monotonic increasing ${\mathcal T}(\varepsilon)$,  the term ${\mathcal T}_0^{\prime}/(1-{\mathcal T}_0)$ has a similar role in the case of a heating spot downstream,  and the opposite one in the case of a cooling spot upstream.  

The experimental observation of a hot spot downstream of a quantum-point-contact,  together with the absence of a cool spot upstream \cite{zeldov_pc2019} is in line with our findings.  However,  other reasons for the absence of a cool spot can be invoked,  like the different scales over which the local temperatures are defined (that are not part of our description where we work at fixed temperature and determine the existence of heating and cooling spots,  rather than hot and cool spots). 

\section{Conclusions}
\label{sec:conclusions}

Dissipation associated with the electronic transport through a coherent obstacle,  {\it i.e.} an elastic scatterer,  has been analyzed by combining the Landauer-Büttiker scattering formalism with a Boltzmann-like description of the electrodes.  The dissipative electrodes have been likened to quasi-one-dimensional wires,  adopting two model approaches in order to describe the inelastic scattering: an energy-independent mean-free-path and an energy-independent relaxation-time.  These two special cases are physically relevant,  and allow for an analytic treatment of the self-consistent problem,  particularly in the weak-excitation,  low-temperature regime.  

Concerning charge transport,  we have obtained the current-voltage characteristics of the scatterer,  generalizing the well-known 4-point Landauer conductance formula by including self-consistent charging effects and dissipation in the electrodes.  The two models (Eqs.~\eqref{eq:j4plT} and \eqref{eq:j4plTtau}) result in equivalent expressions at zero temperature,  and only differ in their low-temperature correction.  The self-consistent treatment allowed to quantify the Landauer resistivity dipole at both sides of the scatterer (Eqs.~\eqref{eq:deltanlT} and \eqref{eq:deltanlrcrt}),  which could be interesting in view of recent developments of nanometer-resolution measurements of local charge accumulation associated with electronic transport \cite{nanolett2025}.

Concerning energy transport,  we have focused on two experimentally relevant features characterizing non-local dissipation at the nano-scale which have received considerable attention in the literature: the asymmetry between the dissipated power at both sides of the obstacle \cite{lee2013},  and the existence of thermal spots,  away from the obstacle,  where the dissipated power presents a local maximum or minimum (detectable through nano-scale temperature profiles \cite{zeldov_pc2019}). 

The asymmetry of the power dissipation,  with an enhancement downstream of the scatterer,  obtained within the scattering approach in the case in which the transmission coefficient is monotonically increasing with energy \cite{SciPostPhys2022,zotti2014},  can be reinforced or diminished according to the different physical parameters within the two models (Eqs.~\eqref{eq:paecmfp},  \eqref{eq:paf},  \eqref{eq:adpeicrt},  and \eqref{eq:paftau}).  Such a result stresses the need to bo beyond the scattering formalism and is in line with the findings of Ref.~\cite{sanchez2015},  although in that approach,  the inelastic processes are not supposed to occur at the electrodes, but within the scatterer. 

The existence of thermal spots is contingent on the energy-dependence of the mean-free-path associated with the inelastic processes in the electrodes,  and thus,  quite different for the two models treated in our study.  On the one hand,  an energy-independent mean-free-path leads to cooling spots resulting from the countering between inelastic and electrostatic characteristic lengths (Eq.~\eqref{eq:hotspotlT}),  that disappear in the strong Coulomb limit of a diverging carrier charge.  On the other hand,  an energy-independent relaxation-rate results in heating (cooling) spots downstream (upstream) of the scatterer,  provided that some restrictions depending on the system parameters are met.  The position of these last spots has been obtained in the strong Coulomb limit (Eq.~\eqref{eq:hotspotlTcrt}),  and its corrections associated with electrostatic effects have been determined (Eq.~\eqref{eq:deltaz}).  Finite temperature and the non-linear voltage regime are crucial in order to obtain heating spots downstream of the scatterer,  in line with the experimental findings dealing with a quantum constriction embedded in a two-dimensional electron gas \cite{zeldov_pc2019},  where the hypothesis of an energy-independent relaxation-rate appears as more realistic than that of an energy-independent mean-free-path.

The voltage drop $V$ induced by the scatterer and the system temperature $T$ are the two critical parameters that define the regime of weak-excitation,  low-temperature that we have analyzed by assuming small values of $eV/\mu_0$ and $k_{\rm B}T/\mu_0$.  In addition,  there are several system parameters which determine a wealth of possible physical behavior.  For metallic samples,  where the present approach captures the essence of the three-dimensional problem  \cite{eranen1987},  some typical parameters are: $n^{(0)}=10^{21}$ ${\rm cm}^{-3}$,  $\mu_{0}=4$ eV,  $k_{\rm s}^{-1} \simeq 1$ \AA, $l \simeq 500$ \AA, $L \simeq 50$ \AA, and $\epsilon_{\rm B} \simeq 2\epsilon$,  while in low-dimensional semiconductor nanostructures, {\it i.e.} quantum wires \cite{PhysRevApp2020,Limpert2017,LinkePRL2024} or hetero-junctions \cite{zeldov_pc2019},  considerably smaller values of $n^{(0)}$ and $\mu_{0}$ should be used,  and a poorer screening is expected.

Instead of selecting plausible values of some physical constants and working conditions,  followed by a numerical exploration of the results as a function of the other variables of the problem,  we have identified the key parameters and gave general expressions in terms of them,  which are  applicable to different scenarios.  According to the value of these key parameters,  we obtain different regimes  and,  in some limiting cases,  simplified expressions result. For instance,  placing ourselves in the case of good screening with $k_{\rm s}l \gg 1$,  as we have done in some chapters,  considerably simplifies the form of the current-voltage characteristics and the power dissipation,  and leads to results which are quite different than those of the opposite limit, above discussed,  where screening is poor.  The thermal factor (Eqs.~\eqref{eq:therfac} and \eqref{eq:therfacwh},  related to the value of the Seebeck coefficient) determines the importance of thermal effects.  Simplified expressions of the thermal corrections are obtained under the condition $\mathcal{T}^{\prime}_0 \gg \mathcal{T}_0/\mu_0$,  which is generically applicable,  except in conductance plateaus where the opposite limit is reached.  The excitation-dependent parameter $\gamma$ (Eqs.~\eqref{eq:defctesgamma},  \eqref{eq:gammalT},  and \eqref{eq:gammalTctau}) sets the strength of the electrostatic modifications with respect to the equilibrium configuration. 

The proposed formalism is restricted to the case of degenerate electrons,  in contrast to the numerical approach of Ref.~\cite{NicoPRB},  establishing the ubiquity of thermal spots in the non-degenerate case.  Moreover,  Ref.~\cite{NicoPRB} worked in the strong Coulomb limit and focused on the case of transmission coefficients acting as energy filters that enhance the thermoelectric effects \cite{svensson2012lineshape,Limpert2017,LinkePRL2024,BENENTI2017},  while our analysis includes the non-homogeneity of the carrier density and it has been developed for cases with a smooth energy-dependence of the scatterer's transmission coefficient.  Ref. ~\cite{Mirlin2019} also worked in the strong Coulomb limit,  identifying power dissipation asymmetries and hot spots in the hydrodynamic regime (where the electron-electron relaxation time is smaller than their impurity and phonon counterparts),  while the inelastic electron-phonon scattering gives rise to the only relaxation mechanism that we considered in this work.   

The essential ingredients determining the power dissipation asymmetry and the existence of thermal spots that we have identified in our model system (finite temperature,  non-linear regime,  as well as the energy dependence of the scatterer's transmission coefficient and of the carriers' mean-free-path)  are expected to also prevail in the generalizations describing experimentally realizable setups.  These ingredients are indeed present in the case studied in Ref. ~\cite{zeldov_pc2019} of a quantum-point-contact defined in a  semi-conductor hetero-structure.  In order to advance in the connection with this kind of experiments,  other than the change in the values of some key parameters and the generalization beyond the quasi-one-dimensional geometry,  we will need to take into account the poor screening within the two-dimensional electron gas \cite{jalabert1989},  and include the relaxation induced by electron-electron interaction, through the Boltzmann equation \cite{Mirlin2019} or from the quasi-particle self-energy \cite{SciPostPhys2022}.  

The proposed way of accounting for dissipation in the context of coherent scattering is complementary to that of the phenomenological Büttiker fictitious leads \citep{Butt_PRB86},  where inelastic effects and phase randomization enter through the coupling of the system to auxiliary reservoirs.  In thermoelectric studies,  such an approach has been implemented by attaching the fictitious leads to the scatterer \cite{sanchez2015} or to the electrodes \cite{GoriniPRB2021}.  Bridging the gap between these two complementary approaches would be important in order to advance in our understanding of dissipation in the  neighborhood of a coherent scatterer.   Alternatively,  dissipation can be quantified by the rate of entropy production \citep{AM,Gurevich97,bringuier_hal2016}.  The approach presented in this work allows to obtain the non-equlibrium distribution functions for the carriers,  and thus the entropy production.  

We have not considered in this work the presence of a magnetic field that results in edge channels.  A possible generalization of our work in this direction would be relevant in order to analyze studies of energy transport \cite{parmentier2013,parmentier2018} and dissipation \citep{marguerite2019,Xie2024} in chiral configurations,  as well as to relate to recent generalizations of the  Landauer scattering approach with distributed injection along the length of the channel \cite{edgePRL2025}.


\section*{Acknowledgments}

We thank financial support from the French National Research Agency ANR through Project No. ANR-20-CE30-0028 (TQT).  D.M. Basko,  N.G.  Leumer,  D. Weinmann,  and R.S. Whitney are gratefully acknowledged for a previous collaboration that led to this work,  as well as for rich discussions on the subject.  We are particularly indebted to D. Weinmann for a careful reading of the manuscript and useful comments. 

\appendix

\section{Self-consistent treatment of the electrostatic potential for a velocity-independent mean-free-path}
\label{sec:appendixSCTEP}

The self-consistent treatment of the electrostatic problem involves the solution of Poisson equation within the wire
\begin{equation}
\label{eq:Poisson}
\nabla^2 \phi_{\rm L,R}(z) = -\frac{4\pi e}{\epsilon} \ \delta n_{\rm L,R}(z) \,  ,
\end{equation}
where $\epsilon$ stands for its permittivity and $e$ is the carriers' charge.  In the case of a velocity-independent mean-free-path $\ell$,  the form \eqref{eq:deltan} of the density correction results in 
\begin{widetext}
\begin{equation}
\label{eq:Poisson2}
\nabla^2 \phi_{\rm L,R}(z) = -\frac{4\pi e}{\epsilon} 
\left\{
-2 \zf 
\left(
\delta \mu_{\rm L,R}(z_{\mp}) + \frac{j}{2 \zj \ell} (z-z_{\mp}) - 
e\phi_{\rm L,R}(z)
\right)
\pm q_{\rm L,R} \ \exp{\left(\pm \ \frac{z-z_{\mp}}{\ell} \right)} 
\right\}
  \,  .
\end{equation}

The electrostatic potential at the left (right) of the scatterer is thus given by  \cite{eranen1987}
\begin{subequations}
\label{eq:elpot}
\begin{align}
\label{eq:elpotL}
\phi_{\rm L}(z) & = A_{\rm L} \ \exp{\left(k_{\rm s}\left[ z-z_{-} \right] \right)} + \frac{j}{2e \zj \ell} \left[ z-z_{-} \right] + 
\frac{\delta \mu_{\rm L}(z_{-})}{e} +
\frac{4\pi e}{\epsilon} \  \frac{\ell^2}{k_{\rm s}^2\ell^2-1} \ q_{\rm L} \
\exp{\left(\frac{z-z_{-}}{\ell} \right)} 
 \,  ,
\\
\phi_{\rm R}(z) & = A_{\rm R} \ \exp{\left(k_{\rm s}\left[ z_{+}-z \right] \right)} + \frac{j}{2e \zj \ell} \left[ z-z_{+} \right] + 
\frac{\delta \mu_{\rm R}(z_{+})}{e} -
\frac{4\pi e}{\epsilon} \  \frac{\ell^2}{k_{\rm s}^2\ell^2-1} \ q_{\rm R} \
\exp{\left(\frac{z_{+}-z}{\ell} \right)} 
 \,  ,
\end{align}
\end{subequations}
\end{widetext}
with
\begin{equation}
\label{eq:isl}
k_{\rm s} = \sqrt{\frac{4\pi e^2}{\epsilon} \ 2 \left| \zf \right|}
\end{equation}
the inverse screening length (having the Thomas-Fermi form of the quasi-one dimensional geometry),  and $\zf$ given by Eq.~\eqref{eq:f0}.

Assuming that the scatterer is electrically neutral and with a permittivity $\epsilon_{\rm B}$,  the electrostatic potential within is simply $\phi_{\rm B}(z) = A_{\rm B} \ z +  B_{\rm B}$.  The matching conditions at the interfaces $z=z_{\mp}$ result in a linear system for the parameters $A_{\rm L}$,  $A_{\rm R}$,  $A_{\rm B}$,  and $B_{\rm B}$ that yields \cite{eranen1987}
\begin{widetext}
\begin{subequations}
\label{eq:ALAR}
\begin{align}
A_{\rm L}  & = - \frac{\gamma}{e}
- \frac{1}{\omega} \
\frac{4\pi e}{\epsilon k_{\rm s}} \ \ell
\left[
q_{\rm L} \left( 
\frac{\epsilon k_{\rm s} L}{k_{\rm s}^2\ell^2-1} + 
\frac{\epsilon_{\rm B}}{k_{\rm s}\ell-1}
\right)
+ q_{\rm R} \ \frac{\epsilon_{\rm B}}{k_{\rm s}\ell+1}
\right]
 \,  ,
\\
A_{\rm R}  & =  \frac{\gamma}{e} +
\frac{1}{\omega} \
\frac{4\pi e}{\epsilon k_{\rm s}} \ \ell
\left[
q_{\rm L} \ \frac{\epsilon_{\rm B}}{k_{\rm s}\ell+1}
+ q_{\rm R} 
\left( 
\frac{\epsilon k_{\rm s} L}{k_{\rm s}^2\ell^2-1} + 
\frac{\epsilon_{\rm B}}{k_{\rm s}\ell-1}
\right)
\right]
 \,  ,
\end{align}
\end{subequations}
 \end{widetext}
with the system-dependent parameter $\omega$ defined in Eq.~\eqref{eq:omega},  and the excitation-dependent parameter given by
\begin{equation}
\label{eq:defctesgamma}
 \gamma  = \frac{1}{\omega} 
 \left(
e V \epsilon_{\rm B} + \frac{j \epsilon L}{2 \zj \ell} 
\right)
 \,  .
\end{equation}
It is straightforward to verify that the previous form of $A_{\rm L,R}$ respects the global particle-conserving condition ${\cal N}_{\rm L}+{\cal N}_{\rm R}= 0$.  

Eq.~\eqref{eq:nuLR} defining the parameters $\nu_{\rm L,R}$,  together with \eqref{eq:elpot} and \eqref{eq:ALAR},  leads to the self-consistency conditions
\begin{subequations}
\label{eq:nuLnuR0}
\begin{align}
\nu_{\rm L}  & =  
-a q_{\rm L} +b q_{\rm R} + \gamma
 \,  ,
\\
\nu_{\rm R}  & = 
-b q_{\rm L} + a q_{\rm R} -\gamma
 \,  ,
\end{align}
\end{subequations}
where we have defined the system-dependent constants
\begin{subequations}
\label{eq:defab}
\begin{align}
\label{eq:defaba}
a & = \frac{4\pi e^2}{\epsilon k_{\rm s}} 
\left( \frac{\ell}{k_{\rm s}\ell+1} \right)
\frac{\epsilon k_{\rm s} L + \epsilon_{\rm B}}{\omega} 
\,  ,
\\
\label{eq:defabb}
b & = \frac{4\pi e^2}{\epsilon k_{\rm s}} 
\left( \frac{\ell}{k_{\rm s}\ell+1} \right)
\frac{\epsilon_{\rm B}}{\omega} 
\,  ,
\end{align}
\end{subequations}
Putting together Eqs.~\eqref{eq:qLR0} and \eqref{eq:nuLnuR0},  results in a linear system for $\nu_{\rm L,R}$ with solutions \cite{eranen1987}
\begin{widetext}
\begin{subequations}
\label{eq:nuLnuR}
\begin{align}
\nu_{\rm L}  & = \frac{1}{\Delta} 
\left\{
\gamma-a\alpha+b\beta
-
\gamma \left[a+b\right]
\left[
 \mathcal{T}_{\rm L}^{\rm R} - 
 \mathcal{T}_{\rm R}^{\rm R}\right]-
\left[a^2-b^2\right]
\left[
\alpha \mathcal{T}_{\rm R}^{\rm R} - 
\beta \mathcal{T}_{\rm L}^{\rm R}\right] 
\right\}
 \,  ,
\\
\nu_{\rm R}  & = \frac{1}{\Delta} 
\left\{
-\gamma-b\alpha+a\beta+
\gamma \left[a+b\right]
\left[
 \mathcal{T}_{\rm R}^{\rm L} - 
 \mathcal{T}_{\rm L}^{\rm L}\right] +
\left[a^2-b^2\right]
\left[
\beta \mathcal{T}_{\rm L}^{\rm L} - 
\alpha \mathcal{T}_{\rm R}^{\rm L}\right] 
\right\}
 \,  ,
\end{align}
\end{subequations}
and
\begin{equation}
\label{eq:defdelta}
\Delta  = 1 +
a \left[
 \mathcal{T}_{\rm L}^{\rm L} +
 \mathcal{T}_{\rm R}^{\rm R}\right] -
b \left[
 \mathcal{T}_{\rm L}^{\rm R} +
 \mathcal{T}_{\rm R}^{\rm L}\right] 
+
\left[a^2-b^2\right]
\left[
 \mathcal{T}_{\rm L}^{\rm L} \
 \mathcal{T}_{\rm R}^{\rm R}
- 
 \mathcal{T}_{\rm R}^{\rm L}
 \mathcal{T}_{\rm L}^{\rm R}
\right] 
\,  .
\end{equation}
 \end{widetext}
The solution \eqref{eq:nuLnuR} of the electrostatic problem yielding $\nu_{\rm L,R}$,  when used in Eq.~\eqref{eq:j4p},  leads to an implicit self-consistent current-voltage characteristics.  An explicit current-voltage characteristics can be obtained by noticing that the form \eqref{eq:alphadef},  \eqref{eq:betadef},  and  \eqref{eq:defctesgamma} of the parameters $\alpha$,  $\beta$,  and $\gamma$,  respectively,  present an obvious separation between $j$ and $V$ components,  allowing to write  
\begin{subequations}
\label{eq:jVcomponentsctes}
\begin{align}
\alpha & =  \frac{j}{\zj} \  \alpha^{(j)} +  \alpha^{(V)}
 \,  ,
\\
\beta & =  \frac{j}{\zj}  \  \beta^{(j)} +  \beta^{(V)}
 \,  ,
\\
\gamma &  =  \frac{j}{\zj} \  \gamma^{(j)} +  \gamma^{(V)}
\,  .
\end{align}
\end{subequations}
with

\begin{subequations}
\label{eq:defctesj}
\begin{align}
\label{eq:alphadefj}
\alpha^{(j)} & = 
 \zf - \frac{1}{2}
 \left[
 \mathcal{T}_{\rm L}^{\rm L} + 
 \mathcal{T}_{\rm L}^{\rm R}
\right]
 \,  ,
\\
\label{eq:betadefj}
\beta^{(j)} & = 
\zf - \frac{1}{2}
 \left[
 \mathcal{T}_{\rm R}^{\rm R} + 
\mathcal{T}_{\rm R}^{\rm L}
\right]
 \,  ,
 \\
\label{eq:gammadefj}
\gamma^{(j)}  &= 
 \frac{\epsilon \ L}{2 \ell \ \omega} 
\,  ,
\end{align}
\end{subequations}
and
\begin{subequations}
\label{eq:defctesV}
\begin{align}
\label{eq:alphadefV}
\alpha^{(V)} & =  - 
 \frac{m}{\pi A \hbar} \int_{0}^{\infty} {\rm d}v_{\rm L} \ 
\mathcal{T}(\varepsilon) 
\left[ f(\ecL) - f(\ecR) \right]
 \,  ,
\\
\label{eq:betadefV}
\beta^{(V)} & = - 
 \frac{m}{\pi A \hbar} \int_{0}^{\infty} {\rm d}v_{\rm R} \ 
\mathcal{T}(\varepsilon) 
\left[ f(\ecL) - f(\ecR) \right]
 \,  ,
 \\
\label{eq:gammadefV}
 \gamma^{(V)} &  = \frac{\epsilon_{\rm B} }{\omega} 
e V 
\,  .
\end{align}
\end{subequations}

Using Eqs.~\eqref{eq:nuLnuR},  we see that  $\nu_{\rm L,R}$ inherit a similar separation as above,  and thus,  we can write
\begin{equation}
\nu_{\rm L,R} =  \frac{j}{\zj} \  \nu_{\rm L,R}^{(j)} +  \nu_{\rm L,R}^{(V)}
\,  ,
\end{equation}
which allows to go from Eq.~\eqref{eq:j4p} to the explicit current-voltage characteristics
 \begin{widetext}
\begin{equation}
\label{eq:j4pex}
j = \frac{1}{1-\frac{1}{2}\left(\tL + \tR \right)+ \nu_{\rm L}^{(j)} \ \tL -  \nu_{\rm R}^{(j)}\ \tR} \left\{  \frac{em}{\pi A \hbar} \int_{0}^{\infty} {\rm d}v_{\rm L} \ v_{\rm L} \
\mathcal{T}(\varepsilon) 
\left[ f_{0}(\ecL) - f_{0}(\ecR)  \right] - \zj \ \nu_{\rm L}^{(V)} \ \tL +
\zj \ \nu_{\rm R}^{(V)} \ \tR \right\} \,  .
\end{equation}
 %
For a generic $\mathcal{T}(\varepsilon)$ the final form of the current-voltage characteristics is obtained after performing the various energy-integrations of the above-defined system parameters.  Further analytical progress can be made by assuming a smooth energy-dependence of $\mathcal{T}(\varepsilon)$,  while working in the weak-excitation and low-temperature limit,  as we do in Appendix \ref{sec:Appltlv}.
\end{widetext}

\section{Weak-excitation,  low-temperature limit for a velocity-independent mean-free-path}
\label{sec:Appltlv}

In the case of a degenerate electron gas fulfilling $k_{\rm B}T/\mu_0 \ll 1$ and a function $\mathcal{T}(\varepsilon)$ that is smoothly-varying in the scale of $k_{\rm B}T$,  we can use a systematic low-temperature Sommerfeld expansion around the equilibrium chemical potential

\begin{equation}
\mu_0 = \EF \left[1+ \frac{\pi^2}{12} \left(\frac{k_{\rm B}T}{ \EF} \right)^2 \right] \,  ,
\end{equation}
where we note $\EF$ the Fermi energy.  To first-order in $\left(k_{\rm B}T/\EF \right)^2$,  the parameters $\zf$ and $\zj$,  defined respectively by Eqs.~\eqref{eq:f0} and \eqref{eq:J0},  become

\begin{subequations}
\begin{align}
\label{eq:f0T0}
\zf & = - \frac{1}{\pi A \hbar v_{0} }
\left[1+ \frac{\pi^2}{8} \left(\frac{k_{\rm B}T}{ \mu_0}\right)^2 \right] \,  ,
\\
\label{eq:j0T0}
\zj & = - \frac{e}{\pi A \hbar} 
\,   .
\end{align}
\end{subequations}
The low-$T$ expansions of the other important parameters are 
\begin{widetext}
\begin{subequations}
\label{eq:expansctestauLRLRT0}
\begin{align}
 \mathcal{T}_{\rm L,R} & =  \tilde{\mathcal T}(\mu_0 \pm e V/2) \,  ,
\\
\mathcal{T}_{\rm L,R}^{\rm L,R} & = - \frac{1}{\pi A \hbar \ v_{0} } \
\Biggl[
 \mathcal{\tilde T}(\mu_0 \pm e V/2) -  \frac{\pi^2}{6} 
 \frac{\left(k_{\rm B}T\right)^2}{ \mu_0} 
\left\{
 \mathcal{T}^{\prime}(\mu_0 \pm e V/2) - \frac{3}{4\mu_0} 
 \mathcal{T}(\mu_0 \pm e V/2)
\right\}
\Biggr]
 \,  \\
\mathcal{T}_{\rm L,R}^{\rm R,L} & = 
- \frac{1}{\pi A \hbar \ v(\mu_{0} \mp e V) } \
\Biggl[
 \mathcal{\tilde T}(\mu_0 \mp e V/2) -  \frac{\pi^2}{6} 
 \frac{\left(k_{\rm B}T\right)^2}{ \mu_0}
\left\{
 \mathcal{T}^{\prime}(\mu_0 \mp e V/2) - \frac{3}{4\mu_0} 
 \mathcal{T}(\mu_0 \mp e V/2)
\right\}
\Biggr]
\,  .
\end{align}
\end{subequations}
We have defined 
\begin{equation}
\label{eq:tautilde}
\tilde{\mathcal T}(\varepsilon)={\mathcal T}(\varepsilon) + \frac{\pi^2}{6}\left(k_{\rm B}T\right)^2 \ {\mathcal T}^{\prime\prime}(\varepsilon) \,  ,
\end{equation}
and we use the sub-index 0 for the value of the energy-dependent functions at $\mu_0$ ({\it i.e. }$v_0=v(\mu_0)$).  The above expressions can be further simplified if we assume that the function $\mathcal{T}(\varepsilon)$ is smoothly-varying in the scale of $e V$.  Working in the weak-excitation limit,  we perform a systematic expansion up to second-order terms in $e V/\mu_0$.  Under such a hypothesis,  the parameter $\Delta$ defined in \eqref{eq:defdelta} can be taken as

\begin{equation}
\label{eq:deltalT}
\Delta  = 1 - \frac{ \epsilon k_{\rm s} L}{\omega} 
\frac{k_{\rm s} \ell}{k_{\rm s}\ell+1}
\left(
\tilde{\mathcal T}_0
-
\frac{\pi^2}{6}  \frac{\left(k_{\rm B}T\right)^2}{\mu_0} 
{\mathcal T}_0^{\prime} 
\right)
\,  ,
\end{equation}
and the explicit current-voltage characteristics \eqref{eq:j4pex} takes the form presented in Eq.~\eqref{eq:j4plT} of the main text.  Using this last result,  together with the definition \eqref{eq:defctesgamma},  we have
\begin{equation}
\label{eq:gammalT}
\gamma = \frac{e V}{2}
\left(\frac{2\epsilon_{\rm B}/\omega-
{\tilde \kappa}}{1+{\tilde \kappa }}  +
\frac{1}{2} \ \frac{{\mathcal T}_0}{1-\mathcal{T}_0 } \ d_{\rm T} \
\frac{ \epsilon k_{\rm s} L}{\omega} \
\frac{\kappa \left(1+2\epsilon_{\rm B}/\omega\right)}{\left(1+\kappa \right)^2} \
\frac{k_{\rm s} \ell}{k_{\rm s}\ell+1}  
\right)
\,  .
\end{equation}
The sign of $\gamma$,  set by the dominant zero-temperature contribution (the first term within the parenthesis),  is determinant in establishing the behavior of various physical quantities of the problem at hand.  Since the scatterer is ballistic,  {\it i.e. } $L/\ell \ll 1$,  we have that $\gamma >0$ under very general conditions.  Indeed,  $\gamma$ is negative only for transmission coefficients so close to unity that $1- \mathcal{T}_0 < L/2l$ (assuming that $\epsilon$ and $\epsilon_{\rm B}$ are of the same order).  In the case of perfect transmission ($ \mathcal{T}_0 \simeq 1$), we approach the homogeneous Drude problem,  where $\gamma \simeq - e V/2$,  while Eq.\eqref{eq:j4plT} yields $-j/2j_0l \simeq 2 e V/L$,  implying that $V$ (and then $\gamma$) scales to 0 with $L$.  

From Eq.~\eqref{eq:j4plT} we also obtain
\begin{align}
\label{eq:nuLnuRlT}
\nu_{\rm L,R}  =  \pm &
\left(
\gamma  -
\frac{e V}{4} \ \frac{{\mathcal T}_0}{1-\mathcal{T}_0 } \ d_{\rm T} \
\frac{ \epsilon k_{\rm s} L}{\omega} \
\frac{1+2\epsilon_{\rm B}/\omega}{1+\kappa} \
\frac{k_{\rm s} \ell}{k_{\rm s}\ell+1}   
\right)
+
 \frac{\pi A \hbar}{8\mu_0}  \  j \  V\frac{k_{\rm s} \ell}{k_{\rm s}\ell+1}
\nonumber
\\
& - (eV)^2\
\frac{\pi^2}{16} \left( \frac{k_{\rm B}T}{ \mu_0}\right)^2 \
\frac{1+2\epsilon_{\rm B}/\omega}{1+\kappa} \
\frac{1}{1-\mathcal{T}_0} 
\left(
{\mathcal T}^{\prime}_0 - \frac{\mathcal{T}_0}{\mu_0}
\right)
 \frac{k_{\rm s} \ell}{k_{\rm s}\ell+1}
 \,  ,
\end{align}
and
\begin{equation}
\label{eq:qLRlT}
q_{\rm L,R } =  
 \frac{\ eV}{2 \pi A \hbar v_{0} }  \ 
 \frac{{\mathcal T}_0}{1-\mathcal{T}_0 } \ d_{\rm T} \
\frac{1+2\epsilon_{\rm B}/\omega}{1+\kappa} 
\mp \frac{ j \ V}{4v_0\mu_0}  
\pm \frac{(eV)^2}{\pi A \hbar v_{0} } 
\frac{\pi^2}{8} \left( \frac{k_{\rm B}T}{ \mu_0}\right)^2 
\frac{1+2\epsilon_{\rm B}/\omega}{1+\kappa} 
\frac{1}{1-\mathcal{T}_0} 
\left(
{\mathcal T}^{\prime}_0 - \frac{5}{4} \ \frac{\mathcal{T}_0}{\mu_0}
\right)
 \,  .
\end{equation}
Similarly,  the parameters $A_{\rm L,R}$ of Eqs. \eqref{eq:elpot}-\eqref{eq:ALAR} become
\begin{align}
\label{eq:ALARlT}
A_{\rm L,R}  =  \mp &
\left(
\frac{\gamma}{e}
 +
 \frac{V}{4} \ \frac{{\mathcal T}_0}{1-\mathcal{T}_0 } \ d_{\rm T} \
\frac{1+2\epsilon_{\rm B}/\omega}{1+\kappa} \
\left[
1 + \frac{2\epsilon_{\rm B}}{\omega} \left( k_{\rm s} \ell - 1 \right) 
\right] \
\frac{k_{\rm s} \ell}{k_{\rm s}^2 \ell^2-1}  
\right)
+ 
\frac{\pi A \hbar}{8\mu_0 e}  \  j \  V
\frac{k_{\rm s} \ell}{k_{\rm s}^2\ell^2-1}
\nonumber \\
&
-  e V^2 \
\frac{\pi^2}{16} \
\left(\frac{k_{\rm B}T}{ \mu_0}\right)^2
\frac{1+2\epsilon_{\rm B}/\omega}{1+\kappa} \
\frac{1}{1-\mathcal{T}_0} 
\left(
{\mathcal T}^{\prime}_0 - \frac{\mathcal{T}_0}{\mu_0}
\right)
 \frac{k_{\rm s} \ell}{k_{\rm s}^2\ell^2-1}
 \,  . 
\end{align}

Using Eq.~\eqref{eq:deltan} and the previous results,  we see that the decay of $\delta n_{\rm L,R}(z)$ away from the scatterer is not simply given by $k_{\rm s}$,  as inferred within a straightforward extension of the non-interacting scattering approach \citep{datta1995},  but also depends on the length scale set by $\ell$.  Moreover,  the density variations at the two sides of the scatterer have the form
\begin{align}
\label{eq:deltanlT}
\delta n_{\rm L,R}(z_{\mp})  =  &
\pm 
 \frac{2}{\pi A \hbar v_{0}}
\left(
\gamma
 +
 \frac{eV}{2} \
\frac{\pi^2}{6} \frac{\left(k_{\rm B}T\right)^2}{ \mu_0} \
\left[
\frac{{\mathcal T}^{\prime}_0}{1-\mathcal{T}_0} \
\frac{1+2\epsilon_{\rm B}/\omega}{1+\kappa} \
\left(
1 - \frac{ \epsilon k_{\rm s} L}{\omega} \
\frac{k_{\rm s} \ell}{k_{\rm s}\ell+1}  \right) 
+
\frac{3}{4\mu_0}
\right]  
\right)
\nonumber \\
&
- 
\frac{1}{4v_{0} \mu_{0}}  \  j \  V \
 \frac{1}{k_{\rm s}\ell+1} 
+
\frac{(eV)^2}{\pi A \hbar v_{0} } \
\frac{\pi^2}{8} \left( \frac{k_{\rm B}T}{ \mu_0}\right)^2 \
\frac{1+2\epsilon_{\rm B}/\omega}{1+\kappa} \
\frac{1}{1-\mathcal{T}_0} 
\left(
{\mathcal T}^{\prime}_0 - \frac{5}{4} \ \frac{\mathcal{T}_0}{\mu_0}
\right)
 \frac{1}{k_{\rm s}\ell+1}
 \,  . 
\end{align}
\end{widetext}
The dominant contribution to $\delta n_{\rm L,R}(z_{\mp})$,  given by $\pm \gamma$,  sets the sign of the charge accumulation at both sides of the scatterer.  As above discussed $\gamma > 0$,   and thus the Landauer dipole is made of an accumulation (depletion) of carriers upstream (downstream) with respect to the current flow.  

In order to simplify the presentation,  we express some key contributions to $\nu_{\rm L,R}$,  $q_{\rm L,R}$,  $A_{\rm L,R}$,  and $\delta n_{\rm L,R}(z_{\mp})$  in terms of the parameter $\gamma$ or the product of $j$ and $V$,  keeping in mind that these three quantities are linearly related by Eqs.~\eqref{eq:j4plT} and \eqref{eq:gammalT}.  One advantage of this presentation is to visualize the simple expressions obtained at zero temperature for the different physical variables.  At finite temperature,  we have to take into account the explicit temperature dependence in \eqref{eq:nuLnuRlT} and \eqref{eq:qLRlT},  as well as the temperature dependence of  $\gamma$ and the product $j \ V$. 
The contribution to $q_{\rm L,R}$ which is linear in the excitation vanishes at zero temperature and has the same sign in the two cases,  while the quadratic contribution is the opposite.  On the contrary,  the linear contribution to $\nu_{\rm L,R}$ has opposite sign in the two  cases, while the quadratic contribution is the same one,  and a similar behavior is obtained for $A_{\rm L,R}$,  and $\delta n_{\rm L,R}(z_{\mp})$.  For completeness,  we provide in Eqs.~\eqref{eq:nuLnuRlT}-\eqref{eq:deltanlT} the lowest temperature correction of the terms that are quadratic in the excitation. However,  these contributions will not be taken into account for the problem of power dissipation,  and thus,  as discussed after Eq.~\eqref{eq:chpotccrr},  in the regime $eV \ll k_{\rm B}T \ll \mu_0$,  only the contributions linear in the excitation are relevant.  

The variation of the electrostatic potential between the two ends of the scatterer is given by
\begin{widetext}
\begin{equation}
\label{eq:potdroplT}
\Delta \phi = \phi_{\rm L}(z_{-}) - \phi_{\rm R}(z_{+}) = V - 
\frac{2 \gamma}{e} - \frac{V}{2} \
\frac{{\mathcal T}_0}{1-\mathcal{T}_0} \ d_{\rm T}
\frac{ \epsilon k_{\rm s} L}{\omega} \
\frac{1+2\epsilon_{\rm B}/\omega}{1+\kappa} \
\frac{k_{\rm s} \ell}{k_{\rm s}\ell+1}  
\,  ,
\end{equation}
\end{widetext}
and thus,  it is reduced with respect to the total potential drop $V$ induced by the scatterer.

In the metallic case with a screening length much smaller than the mean-free-path,  we have $k_{\rm s} \ell \gg 1$.  Therefore,  the expressions \eqref{eq:j4plT} and  \eqref{eq:gammalT}-\eqref{eq:potdroplT} become considerably simpler.  

\section{Strong Coulomb limit for a velocity-independent mean-free-path}
\label{sec:scl}

The strong Coulomb limit is obtained for a diverging value of the charge carrier $e$,  and a vanishing of the screening length $ k_{\rm s, }$ such that Eq.~\eqref{eq:isl} remains valid.  Adopting the strong Coulomb limit allows for important simplifications in the treatment of the Boltzmann equation \cite{NicoPRB},  which result in an analytical formulation for the case of a velocity-independent mean-free-path.  In this appendix,  we treat the strong Coulomb limit as a particular case of the previous analysis,  focusing on the effects arising from a finite Coulomb interaction.  

Within the above-defined assumption,  the system-dependent constants defined in Eq.~\eqref{eq:defab} have the limiting values
\begin{subequations}
\label{eq:ctesscl}
\begin{align}
a & \rightarrow - \frac{1}{2 \zf } 
\,  ,
\\
b & \rightarrow 0
\,  ,
\end{align}
\end{subequations}
while the excitation-dependent parameter $\gamma$ introduced in 
\eqref{eq:defctesgamma} has the limit
\begin{equation}
\label{eq:gammascl}
\gamma   \rightarrow 0
 \,  .
\end{equation}
Similarly,
\begin{equation}
\label{eq:ALRscl}
A_{\rm L,R}    \rightarrow 0
 \,  ,
\end{equation}
and
\begin{equation}
\label{eq:deltanscl}
\delta n_{\rm L,R}(z)   \rightarrow 0
 \,  ,
\end{equation}
indicating that the carrier density is uniform along the wire.  The local charge neutrality is respected everywhere in the system,  and in particular,  there is no Landauer dipole (unlike the case described in Fig.~\ref{fig:setup}).  The product $e \phi_{\rm L,R}(z)$ remains finite,  such that

\begin{equation}
\label{eq:ephiscl}
e \phi_{\rm L,R}(z) = \delta \mu_{\rm L,R}(z) \mp 
\frac{q_{\rm L,R}}{2 \ {\cal F}_0} \
 \exp{\left(\pm \ \frac{z-z_{\mp}}{\ell} \right)} 
 \,  ,
\end{equation}
and in particular,  $q_{\rm L,R} =\pm 2 \ \zf \ \nu_{\rm L,R}$.  In the strong Coulomb limit,  the expressions \eqref{eq:nuLnuR} simplify to

\begin{subequations}
\label{eq:nuLnuRscl}
\begin{align}
\nu_{\rm L}  & = \frac{1}{2 \zf \Delta} 
\left\{\alpha- \frac{1}{2 \zf } 
\left[
\alpha \mathcal{T}_{\rm R}^{\rm R} - 
\beta \mathcal{T}_{\rm L}^{\rm R}\right]
\right\}
 \,  ,
\\
\nu_{\rm R}  & = - \frac{1}{2 \zf \Delta} 
\left\{\beta- \frac{1}{2 \zf } 
\left[
\beta \mathcal{T}_{\rm L}^{\rm L} - 
\alpha \mathcal{T}_{\rm R}^{\rm L}\right] 
\right\}
 \,  ,
\end{align}
\end{subequations}
with

\begin{equation}
\Delta  =  1 -  \frac{1}{2 \zf } 
\left[
 \mathcal{T}_{\rm L}^{\rm L} +
 \mathcal{T}_{\rm R}^{\rm R}\right] +
 \frac{1}{4 \zf^2 } 
\left[
 \mathcal{T}_{\rm L}^{\rm L} \
 \mathcal{T}_{\rm R}^{\rm R}
- 
 \mathcal{T}_{\rm R}^{\rm L}
 \mathcal{T}_{\rm L}^{\rm R}
\right] \,  ,
\end{equation}
while $\alpha$ and $\beta$ are defined in Eqs.~\eqref{eq:defctes}.  Using the expressions \eqref{eq:nuLnuRscl} of $\nu_{\rm L,R}$,  the implicit current-voltage characteristics \eqref{eq:j4p} can be written in an explicit form as  
\begin{widetext}
\begin{align}
\label{eq:j4pscl}
j = & \frac{em}{\pi A \hbar}  
 \left\{ \int_{0}^{\infty} {\rm d}v_{\rm L} \ v_{\rm L} \
\mathcal{T}(\varepsilon) 
\left[ f(\ecL) - f(\ecR)  \right] - \right.
\frac{ \zj}{2 \zf e \Delta} \ 
\left(
  \tL - \frac{1}{2 \zf } 
  \left[ \tL \mathcal{T}_{\rm R}^{\rm R} -
  \tR \mathcal{T}_{\rm L}^{\rm L}
  \right]
\right) 
\int_{0}^{\infty} {\rm d}v_{\rm L} \ 
\mathcal{T}(\varepsilon) 
\left[ f(\ecL) - f(\ecR)  \right]
\nonumber
\\
& +
 \left.
 \frac{\zj}{2 \zf e \Delta} \ 
\left(
  \tR - \frac{1}{2 \zf } 
  \left[ \tR \mathcal{T}_{\rm L}^{\rm L} -
  \tL \mathcal{T}_{\rm L}^{\rm R}
  \right]
\right) 
\int_{0}^{\infty} {\rm d}v_{\rm R} \ 
\mathcal{T}(\varepsilon) 
\left[ f(\ecL) - f(\ecR)  \right]
\right\} \,  .
\end{align}
\end{widetext}

\subsection{Weak-excitation,  low-temperature regime}
\label{subsec:weltsclcl}

If,  in addition to the strong Coulomb limit,  we consider the regime of weak-excitation and low-temperature,  the current-voltage characteristics 
\begin{equation}
\label{eq:j4plTscl}
j =  \frac{G_0}{A} \ V
\left( \frac{\mathcal{\tilde T}_0}{1-\mathcal{\tilde T}_0} \right) 
\left(1 -  \frac{1}{2} \ \frac{{\mathcal T}_0}{1-\mathcal{T}_0 } \
d_{\rm T}
\right)
\,  ,
\end{equation}
follows from \eqref{eq:j4pscl},  or alternatively,  by using the limits \eqref{eq:ctesscl} in the general expression of Eq.~\eqref{eq:j4plT}.  Similarly, 
\begin{equation}
\label{eq:nuLnuRlTscl}
\nu_{\rm L,R}  =  
\mp
\frac{e V}{4} \ \frac{{\mathcal T}_0}{1-\mathcal{T}_0 } \
d_{\rm T}
+
 \frac{\pi A \hbar}{8\mu_0}  \  j \  V
 \,  ,
 \end{equation}
 and
\begin{equation}
\label{eq:potdroplTscl}
\Delta \phi = V \left(1 -
\frac{1}{2} \ \frac{{\mathcal T}_0}{1-\mathcal{T}_0 } \ d_{\rm T}
\right)
\,  ,
\end{equation}
indicating that the variation of the electrostatic potential between the two ends of the scatterer coincides,  up to a finite-temperature correction,  with the total potential drop induced by the scatterer. 

\subsection{Power dissipation}
\label{subsec:pdweltsclcl}

In the strong Coulomb limit,  the dissipated power-density component arising from the energy-flow inhomogenity \eqref{eq:pLRecmfp},  and the corresponding asymmetry  \eqref{eq:paecmfp},  respectively, become
\begin{widetext}
\begin{equation}
 \mathcal{P}_{\rm L,R}^{\rm u}  = \frac{1}{2} \ A \ j  \ V 
 \mp 
\frac{e V}{\pi \hbar} \ \frac{1}{1-\mathcal{T}_0}
\left(
 \frac{\pi^2}{3} \left(k_{\rm B}T \right)^2
\mathcal{T}^{\prime}_0 +
\frac{(eV)^2}{4} 
 \left\{
 \frac{ \mathcal{T}^{\prime}_0(1+2\mathcal{T}_0)}{3} +  \frac{\mathcal{T}_0^2}{2 \mu_0}
\right\}  
\right)
 \,  ,
\end{equation}
and
\begin{equation}
\label{eq:paecmfpscl}
\mathcal{P}_{\rm A}^{\rm u} =
\frac{e V}{\pi \hbar} \ \frac{1}{1-\mathcal{T}_0}
\left(
 \frac{2\pi^2}{3} \left(k_{\rm B}T \right)^2
\mathcal{T}^{\prime}_0 +
\frac{(eV)^2}{2} 
 \left\{
 \frac{ \mathcal{T}^{\prime}_0(1+2\mathcal{T}_0)}{3} +  \frac{\mathcal{T}_0^2}{2 \mu_0}
\right\}  
\right)
  \,  .
\end{equation}
Similarly,  from Eq.~\eqref{eq:paf},  we obtain the field component of the asymmetry
\begin{equation}
\mathcal{P}_{\rm A}^{\rm f}  =  -  
\frac{\left(e V\right)^3}{4\pi \hbar \mu_0} \
\left( \frac{\mathcal{T}_0}{1-\mathcal{T}_0} \right)^2
\,  ,
\end{equation}
yielding the total asymmetry of the dissipation power
\begin{equation}
\label{eq:paecmfpscltot}
\mathcal{P}_{\rm A} =
\frac{e V}{\pi \hbar} \ \frac{1}{1-\mathcal{T}_0}
\left(
 \frac{2\pi^2}{3} \left(k_{\rm B}T \right)^2
\mathcal{T}^{\prime}_0 +
\frac{(eV)^2}{2} 
 \left\{
 \frac{ \mathcal{T}^{\prime}_0(1+2\mathcal{T}_0)}{3} - 
 \frac{1}{2 \mu_0} \frac{\mathcal{T}_0^3}{1-\mathcal{T}_0}
\right\}  
\right)
  \,  .
\end{equation}
\end{widetext}
As discussed in Sec.~\ref{subsec:sadpltl},  the positive asymmetry obtained from $\mathcal{P}_{\rm A}^{\rm u}$ in the case where $\mathcal{T}(\varepsilon)$ is a monotonic increasing function is countered by the field-contribution $\mathcal{P}_{\rm A}^{\rm f}$.  

As indicated in \eqref{eq:ALRscl},  the parameters $A_{\rm L,R}$ vanish in the strong Coulomb limit,  while $A_{\rm L,R} k_{\rm s}^2 \ell^2$ scales with $k_{\rm s} \ell$.  However,  according to Eqs. \eqref{eq:extreme} and \eqref{eq:hotspotlT},  this last divergence is not enough to result in a thermal spot at a finite distance from the scatterer.  The non existence of thermal spots in the strong Coulomb limit with a velocity-independent mean-free-path is a consequence of taking a vanishing screening length and thus having $\ell$ as the only length scale of the problem.

\section{Weak-excitation,  low-temperature regime for a velocity-independent relaxation-time}
\label{sec:AppWELTCRL}

In order to simplify the form of Eq.~\eqref{eq:jsccrr},  valid in the weak-excitation,  low-temperature regime with $l_0=v_0 \tau$,  we define the ordinary functionals
\begin{subequations}
\label{eq:functionalAs}
\begin{align}
\label{eq:functionalAL}
{\cal A}_{\rm L}(u) & = 
\int_{-\infty}^{z_{-}} {\rm d}z \ u(z) \
\exp{\left(\frac{z-z_{-}}{l_{0}} \right)} 
 \,  ,
\\
\label{eq:functionalAR}
{\cal A}_{\rm R}(u) & = 
\int_{z_{+}}^{\infty} {\rm d}z \ u(z) \
\exp{\left(\frac{z_{+}-z}{l_{0}} \right)} 
 \,  ,
\end{align}
\end{subequations}
and
\begin{subequations}
\label{eq:functionalBs}
\begin{align}
\label{eq:functionalBL}
{\cal B}_{\rm L}(u) & = 
\int_{-\infty}^{z_{-}} {\rm d}z \ u(z) \
\exp{\left(\frac{z-z_{-}}{l_{0}} \right)} \
\left[ \frac{z-z_{-}}{l_{0}} \right] 
\,  ,
\\
\label{eq:functionalBR}
{\cal B}_{\rm R}(u) & = 
\int_{z_{+}}^{\infty} {\rm d}z \ u(z) \
\exp{\left(\frac{z_{+}-z}{l_{0}} \right)} \
\left[ \frac{z_{+}-z}{l_{0}} \right] 
\,  ,
\end{align}
\end{subequations}
as well as the $z$-dependent two-function functionals
%
\begin{equation}
\label{eq:functionalCder} 
{\mathcal C}^{\prime}_{\rm L}(u_{1},u_{2};z) = 
\frac{\rm d}{{\rm d} z} \
{\mathcal C}_{\rm L}(u_{1},u_{2};z) 
 \,  ,
\end{equation}
and

\begin{widetext}
\begin{align}
\label{eq:functionalCdef} 
& {\mathcal C}_{\rm L}(u_{1},u_{2};z) =
- \frac{1}{4 \mu_0}
\Biggl\{
\int_{-\infty}^{z} {\rm d}z^{\prime} \ u_{1}(z') \
\exp{\left(\frac{z^{\prime}-z}{l_{0}} \right)} 
\left[ \frac{z^{\prime}-z}{l_{0}} \right] 
\left[ \frac{z^{\prime}-z}{l_{0}} + 3 \right]
\nonumber \\
&
+
 \int_{z}^{z_{-}} {\rm d}z^{\prime} \ u_{1}(z') \
\exp{\left(\frac{z-z^{\prime}}{l_{0}} \right)} 
\left[ \frac{z-z^{\prime}}{l_{0}} \right] 
\left[ \frac{z-z^{\prime}}{l_{0}} + 3 \right]
\nonumber \\
& 
-
\int_{-\infty}^{z_{-}} {\rm d}z' \ u_{1}(z') \
\exp{\left(\frac{z+z'-2z_{-}}{l_{0}} \right)} 
\left[ \frac{z+z^{\prime}-2z_{-}}{l_{0}} \right] 
\left[ \frac{z+z^{\prime}-2z_{-}}{l_{0}} + 3 \right]
\Biggr\}
 +
\exp{\left(\frac{z-z_{-}}{l_{0}} \right)} \
\left[ \frac{z-z_{-}}{l_{0}} \right] \
\times
\nonumber \\
& 
\Biggl\{
\Biggl(
- eV - \nu_{\rm L} + \nu_{\rm R}  + 
{\cal A}_{\rm L}(u_1) + {\cal A}_{\rm R}(u_2) 
\Biggr)
\Biggl({\mathcal T}_0^{\prime} - \frac{{\mathcal T}_0}{4 \mu_0} 
\left[\frac{z-z_{-}}{l_{0}} + 3 \right]
\Biggr)
- \frac{{\mathcal T}_0}{2 \mu_0} 
\Biggl( 
{\cal B}_{\rm L}(u_1) + {\cal B}_{\rm R}(u_2) 
\Biggr)
\nonumber \\
& 
+
eV \frac{ j}{\zj} \
\Biggl[
\frac{{\mathcal T}^{\prime}_0}{{\mathcal T}_0} 
\left[
\frac{z-z_{-}}{l_{0}} + 3
\right]
-
{\mathcal T}^{\prime}_0
  + 
\frac{{\mathcal T}_0}{2 \mu_0} 
\left[
\frac{z-z_{-}}{l_{0}} + \frac{7}{2}
\right]
+
\frac{1}{4 \mu_0} 
\biggl( 
\left[\frac{z-z_{-}}{l_{0}} \right]^2 + 
9 \left[ \frac{z-z_{-}}{l_{0}} \right] + 15 
\biggr)
\Biggr]
\Biggr\} \, .
\end{align}
\end{widetext}

As explained in Sec.  \ref{subsec:crt} of the main text,  the solution of Eq.~\eqref{eq:jsccrr},  when using the appropriate approximations for the above-defined functionals,  leads to the form \eqref{eq:chpotccrr} of the electro-chemical potentials in the wires (valid for a smooth energy-dependence of $\mathcal{T}(\varepsilon)$ and without considering the temperature correction of the quadratic contribution in the excitation). 

As discussed after Eq.~\eqref{eq:g1LRcT},  for a velocity-independent relaxation-time we generically have $n_{\rm L,R}(z)=n^{(\rm le)}_{\rm L,R}(z)$,  and therefore $\delta n_{\rm L,R}(z)$ is given by the right-hand-side of  \eqref{eq:deltanle}.  In addition,  when restricted to the WELT regime,  the form \eqref{eq:chpotccrr} of the electro-chemical potential  sets the Poisson equation \eqref{eq:Poisson} to
\begin{widetext}
\begin{align}
\label{eq:Poisson3}
\nabla^2 \phi_{\rm L,R}(z)  = & -\frac{4\pi e}{\epsilon} 
\Biggl\{
-2 \zf 
\biggl(
\delta \mu_{\rm L,R}(z_{\mp}) 
\pm
\frac{j}{2 {\mathcal J}_0}
\left[ {\hat d}_{\rm T}
\mp \frac{eV}{2 \mu_0}
\right]
+
\frac{j}{2 \zj l_{0}} 
\left[
1+\frac{\pi^2}{24}  
\left(\frac{k_{\rm B}T}{\mu_0}\right)^2 
\right]
(z-z_{\mp}) 
- e\phi_{\rm L,R}(z)
\biggr)
\nonumber
\\
&
\mp 
\frac{1}{\pi A \hbar v_0} \
\frac{j}{ {\mathcal J}_0}
\left[{\hat d}_{\rm T}
\mp \frac{eV}{2 \mu_0}
\right]
 \exp{\left(\pm \ \frac{z-z_{\mp}}{l_0} \right)} 
\Biggr\}
 \,  ,
\end{align}
%
where $\zf$ and ${\hat d}_{\rm T}$ are defined by Eqs.~\eqref{eq:f0} and \eqref{eq:therfacwh},  respectively.  The structure of \eqref{eq:Poisson3} is the same as that of \eqref{eq:Poisson2},  up to the identifications 
\begin{subequations}
\label{eq:identifications}
\begin{align}
\delta \mu_{\rm L,R}(z_{\mp})   &  \rightarrowtail  
\delta \mu_{\rm L,R}(z_{\mp}) 
\pm
\frac{j}{2 {\mathcal J}_0}
\left[ {\hat d}_{\rm T}
\mp \frac{eV}{2 \mu_0}
\right]
 \,  ,
\\
\frac{j}{2 \zj \ell }  &  \rightarrowtail  
\frac{j}{2 \zj l_{0}} 
\left[
1+\frac{\pi^2}{24}  
\left(\frac{k_{\rm B}T}{\mu_0}\right)^2 
\right]
 \,  ,
 \\
q_{\rm L,R} 
&  \rightarrowtail  -
\frac{1}{\pi A \hbar v_0} \
\frac{j}{{\mathcal J}_0}
\left[ {\hat d}_{\rm T}
\mp \frac{eV}{2 \mu_0}
\right]
\, .
\end{align}
\end{subequations}
Thus, the resulting electrostatic potential $\phi_{\rm L,R}(z)$ has the form \eqref{eq:elpot} with the parameters $A_{\rm L,R}$ given by Eq.~\eqref{eq:ALAR},  up to the above-mentioned identifications.  As a consequence,  the latter parameters can be expressed as
\begin{equation}
\label{eq:ALARcrt}
A_{\rm L,R}  =  \mp 
\left(
\frac{\gamma}{e}
 -
 \frac{V}{2} \
\frac{1+2\epsilon_{\rm B}/\omega}{1+\kappa} 
\left[
{\hat d}_{\rm T} \
\frac{{\mathcal T}_0}{1-\mathcal{T}_0} 
\left(
\frac{k_{\rm s} l_0}{k_{\rm s}^2 l_0^2-1} -
\frac{2\epsilon_{\rm B}/\omega}{k_{\rm s} l_0 + 1}
\right) 
-
\frac{\pi^2}{24} \left(\frac{k_{\rm B}T}{ \mu_0}\right)^2 
\kappa
\right]  
\right)
+ 
\frac{\pi A \hbar}{4\mu_0 e}  \  j \  V
\frac{k_{\rm s} l_0}{k_{\rm s}^2\l_0^2-1}
 \,  ,
\end{equation}
while,  according to the definition \eqref{eq:nuLR},  we have
\begin{equation}
\label{eq:nuLnuRcrt}
\nu_{\rm L,R}  =  \pm 
\left(
\gamma
 +
 \frac{V}{2} \
\frac{\kappa(1+2\epsilon_{\rm B}/\omega)}{1+\kappa} 
\left[{\hat d}_{\rm T} \
\frac{k_{\rm s}\l_0}{k_{\rm s} l_0 + 1}
-
\frac{\pi^2}{24} \left(\frac{k_{\rm B}T}{ \mu_0}\right)^2 
\right] 
\right)
- 
\frac{\pi A \hbar}{4\mu_0 }  \  j \  V
\frac{1}{k_{\rm s}\l_0+1}
 \,  .
\end{equation}

Using in Eq.~\eqref{eq:jsc} the form \eqref{eq:chpotccrr} of 
$\delta \mu_{\rm L,R}^{\prime}(z)$,  and that of \eqref{eq:nuLnuRcrt} for $\nu_{\rm L,R}$,  leads to the explicit current-voltage characteristics presented in Eq.~\eqref{eq:j4plTtau} of the main text,  while the excitation-dependent parameter $\gamma$ (defined by \eqref{eq:defctesgamma} by using $l_0$ instead of $\ell$) is given by
\begin{equation}
\label{eq:gammalTctau}
\gamma = \frac{e V}{2}
\left(\frac{2\epsilon_{\rm B}/\omega-
{\tilde \kappa}}{1+{\tilde \kappa }}  -
\frac{\kappa \left(1+2\epsilon_{\rm B}/\omega\right)}{\left(1+\kappa \right)^2} \
\left[{\hat d}_{\rm T} \
\kappa \
 \frac{k_{\rm s}l_0}{k_{\rm s}l_0+1}
 -
 \frac{\pi^2}{24} \left(\frac{k_{\rm B}T}{ \mu_0} \right)^2
 \left(
  \kappa - 2  {\mathcal T}_0
 \right) 
  \right] 
\right)
\,  .
\end{equation}

Using Eq.~\eqref{eq:deltanle} and the previous results,  the density variations at the two sides of the scatterer are given by
\begin{align}
\label{eq:deltanlrcrt}
\delta n_{\rm L,R}(z_{\mp})  =  &
\pm 
 \frac{2}{\pi A \hbar v_{0}}
\left(
\gamma
 +
 \frac{eV}{2} \
\frac{1}{1+\kappa} \
\left[ {\hat d}_{\rm T} \ \kappa \
 \frac{k_{\rm s}l_0}{k_{\rm s}l_0+1} \ 
 \left(1+\frac{2\epsilon_{\rm B}}{\omega}\right)
 +
 \frac{\pi^2}{6} \left(\frac{k_{\rm B}T}{ \mu_0}\right)^2 \
 \left(
  \frac{\epsilon_{\rm B}}{2 \omega} \left(3 - \kappa \right)
  - \kappa
 \right) 
  \right]  
\right)
\nonumber \\
&
- 
\frac{1}{2v_{0} \mu_{0}}  \  j \  V \
 \frac{1}{k_{\rm s}\l_0+1} 
 \,  . 
\end{align}
As in the case of a velocity-independent mean-free-path, the dominant contribution to $\delta n_{\rm L,R}(z_{\mp})$,  given by $\pm \gamma$,  sets the sign of the charge accumulation at both sides of the scatterer.  

As in Appendix \ref{sec:Appltlv},  we express some key contributions to $\nu_{\rm L,R}$,  $A_{\rm L,R}$,  and $\delta n_{\rm L,R}(z_{\mp})$ in terms of the parameter $\gamma$ or the product of $j$ and $V$,  keeping in mind that these three parameters are linearly related by Eqs.~\eqref{eq:j4plTtau} and \eqref{eq:gammalTctau}.  Similarly to the case of a velocity-independent mean-free-path,  the contributions to $\nu_{\rm L,R}$, $A_{\rm L,R}$,  and $\delta n_{\rm L,R}(z_{\mp})$ which are linear in the excitation have an opposite sign for the two left (L) and right (R) wires,  while the quadratic contribution is the same one.  

The variation of the electrostatic potential between the two ends of the scatterer is given by
\begin{equation}
\label{eq:potdroplrcrt}
\Delta \phi = \phi_{\rm L}(z_{-}) - \phi_{\rm R}(z_{+}) = V  - 
\frac{2 \gamma}{e} - V \
\frac{1+2\epsilon_{\rm B}/\omega}{1+\kappa} 
\left[{\hat d}_{\rm T} \
\frac{{\mathcal T}_0}{1-\mathcal{T}_0} \
\frac{2\epsilon_{\rm B}/\omega-1}{k_{\rm s} l_0 + 1}
-
\frac{\pi^2}{24} \left(\frac{k_{\rm B}T}{ \mu_0}\right)^2 \ \kappa 
\right] 
\,  ,
\end{equation}
\end{widetext}
and thus,  it is reduced with respect to the total potential drop $V$ induced by the scatterer.

\section{Strong Coulomb limit for a velocity-independent relaxation-time}
\label{sec:sclcrt}

In the strong Coulomb limit with a velocity-independent relaxation-time,  the limiting values \eqref{eq:ctesscl}-\eqref{eq:deltanscl} are also valid,  while
\begin{equation}
\label{eq:nuLnuRcrtscl}
\nu_{\rm L,R}  \rightarrow 0  
 \,  ,
 \end{equation}
 and
\begin{equation}
\label{eq:potdropcrtscl}
\Delta \phi = V 
\,  .
\end{equation}
The current-voltage characteristics \eqref{eq:j4plTtau} reduces to

\begin{equation}
\label{eq:j4ptauscl}
j =  \frac{G_0}{A} \ V
\left( \frac{\mathcal{\tilde T}_0}{1-\mathcal{\tilde T}_0} \right) 
\left(1 +  \frac{\pi^2}{12} \left( \frac{k_{\rm B}T}{ \mu_0} \right)^2
{\mathcal T}_0
\right)
\,  ,
\end{equation}
whose only temperature dependence is a small Sommerfeld-like correction.  Unlike the corresponding expression \eqref{eq:j4plTscl} for the case of a velocity-independent mean-free-path in the strong Coulomb limit,  there is no dependence of Eq.~\eqref{eq:j4ptauscl} on $\mathcal{\tilde T}_0^{\prime}$,  {\it i.e.} on the monotonic properties of the function $\mathcal{T}(\varepsilon)$.

The dissipated power-density component arising from the energy-flow inhomogenity \eqref{eq:dpeicrt},  and the corresponding asymmetry  \eqref{eq:adpeicrt},  respectively, become
\begin{widetext}
\begin{equation}
 \mathcal{P}_{\rm L,R}^{\rm u}  = \frac{1}{2} \ A \ j  \ V 
 \mp 
\frac{e V}{\pi \hbar} \ \frac{1}{1-\mathcal{T}_0}
\left(
 \frac{\pi^2}{3} \left(k_{\rm B}T \right)^2
  \left\{
\mathcal{T}^{\prime}_0 -
\frac{\mathcal{T}_0}{2 \mu_0}
\left(
\frac{1}{2} - \mathcal{T}_0
\right)
\right\} 
 +
\frac{(eV)^2}{12} 
 \left\{
 1+2\mathcal{T}_0
\right\}  
\right)
 \,  ,
\end{equation}
and
\begin{equation}
\label{eq:paecmfpscrt}
\mathcal{P}_{\rm A}^{\rm u} =
\frac{e V}{\pi \hbar} \ \frac{1}{1-\mathcal{T}_0}
\left(
 \frac{2 \pi^2}{3} \left(k_{\rm B}T \right)^2
  \left\{
\mathcal{T}^{\prime}_0 -
\frac{\mathcal{T}_0}{2 \mu_0}
\left(
\frac{1}{2} - \mathcal{T}_0
\right)
\right\} 
 +
\frac{(eV)^2}{6} 
 \left\{
 1+2\mathcal{T}_0
\right\}  
\right)
  \,  .
\end{equation}

In the strong Coulomb limit, the conditions \eqref{eq:condhscrt} for the appearance of a thermal spot are considerably simplified,  leading to
\begin{equation}
\label{eq:hotspotlTcrt}
z_{\rm L,R}^{{\rm scl}} = z_{\mp} \mp l_0 \
\left\{
\left(
\frac{eV}{2\mu_0} \ \frac{\mathcal{T}_0}{1-\mathcal{T}_0} \pm 1
\right)
\left(
\frac{6 \ eV \mu_0}{\pi^2 \left(k_{\rm B}T \right)^2} \mp
\frac{4 \mu_0 \mathcal{T}_0^{\prime}}{\mathcal{T}_0}
\right) + 3 - 2 \mathcal{T}_0
\left(
1+ \frac{eV}{2\mu_0}
\right)
\right\} 
 \,  .
\end{equation}

\end{widetext}

The above expressions are only meaningful in the cases where
$z_{\rm L}^{{\rm scl}} < z_{-}$ or $z_{\rm R}^{{\rm scl}} > z_{+}$,  {\it i.e.} when the curly brackets are positive,  thus yielding restrictions on the parameter regime for observing thermal spots at a distance of the order of $l_0$ from the scattarer.  When these conditions are met,  the resulting sign of ${\rm d}^2 p_{\rm L,R}^{\rm s}(z)/{\rm d} z^2 |_{z=z_{\rm L,R}^{{\rm scl}}} $ indicates that there is a cooling (heating) spot at the left (right) of the scatterer.  Transmissions close to unitary,  large voltage drops,  and large values of $\mu_0 \mathcal{T}_0^{\prime}/{\mathcal{T}_0}$ favor the appearance of heating spots,  while the previous factors are in competition concerning the appearance of a cooling spot.   Eq.~\eqref{eq:hotspotlTcrt} also shows the crucial role played by finite temperature,  since for a vanishing temperature the thermal spots are pushed to an infinite distance of the scatterer. 



\bibliography{reffin}

\end{document}